\documentclass[sigconf]{acmart}

\AtBeginDocument{%
  }

\title{\System{}: An Open, Extensible, and Usable Interface for AI Interaction}

\author{Jaeryang Baek}
\email{jaeryang_baek@sfu.ca}
\affiliation{%
  \institution{Simon Fraser University}
  \city{Burnaby}
  \state{BC}
  \country{Canada}
}

\author{Ayana Hussain}
\email{ayana_hussain@sfu.ca}
\affiliation{%
  \institution{Simon Fraser University}
  \city{Burnaby}
  \state{BC}
  \country{Canada}
}

\author{Danny Liu}
\email{danny_liu_5@sfu.ca}
\affiliation{%
  \institution{Simon Fraser University}
  \city{Burnaby}
  \state{BC}
  \country{Canada}
}

\author{Nicholas Vincent}
\email{nvincent@sfu.ca}
\affiliation{%
  \institution{Simon Fraser University}
  \city{Burnaby}
  \state{BC}
  \country{Canada}
}

\author{Lawrence H. Kim}
\email{lawkim@sfu.ca}
\affiliation{%
  \institution{Simon Fraser University}
  \city{Burnaby}
  \state{BC}
  \country{Canada}
}

\renewcommand{\shortauthors}{Baek et al.}

\newcommand{\System}{Open WebUI}
\newcommand{\SystemSpace}{Open WebUI }

\begin{document}

\begin{abstract}
While LLMs enable a range of AI applications, interacting with multiple models and customizing workflows can be challenging, and existing LLM interfaces offer limited support for collaborative extension or real-world evaluation. In this work, we present an interface toolkit for LLMs designed to be open (open-source and local), extensible (plugin support and users can interact with multiple models), and usable. The extensibility is enabled through a two-pronged plugin architecture and a community platform for sharing, importing, and adapting extensions. To evaluate the system, we analyzed organic engagement through social platforms, conducted a user survey, and provided notable examples of the toolkit in the wild. Through studying how users engage with and extend the toolkit, we show how extensible, open LLM interfaces provide both functional and social value, and highlight opportunities for future HCI work on designing LLM toolkit platforms and shaping local LLM-user interaction.
\end{abstract}

\begin{teaserfigure}\centering \includegraphics[width=1\columnwidth]{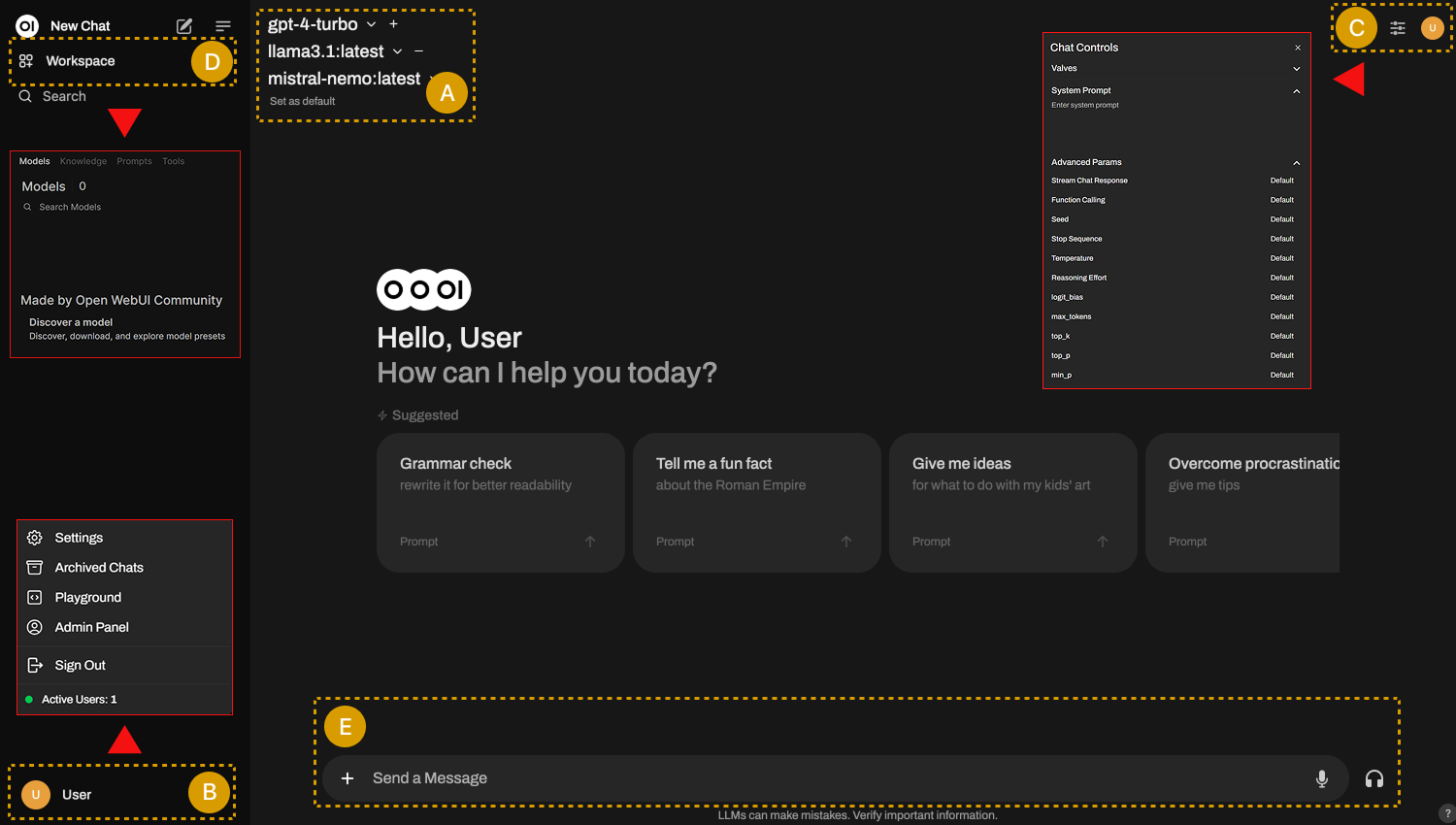} \caption{Screenshot of \System{}. A) The Model Selector, central to the system's design, supports simultaneous interactions with multiple models, including private models (e.g., GPT-4) via APIs. It also allows for directly downloading local models by searching within the selector. B) The \System{} accommodates both standalone and multi-account configurations. C) Users can access the settings menu to manage models. Additional options to adjust model parameters such as seed and temperature are provided through a controls button. D) The sidebar acts as a portal for integrating user-specific models, preset prompts, and documents via Workspace. E) The chat input section is equipped with document, image attachment, web search, and custom tool capabilities via the plus icon; additionally, options for voice and video calls are integrated.} \label{fig:example} \end{teaserfigure}

\maketitle

\section{Introduction}
Sparked by the release of ChatGPT \cite{ray2023chatgpt,huChatGPTSetsRecord2023} in 2022, many subfields of computing -- including social computing \cite{kulkarni2023llms,park2023generative} -- have seen a surge of interest in working on and with large language models (LLMs) \cite{zhao2023survey}. Some scholars have argued that the new generation of ``foundation models'' marks a major shift in computing, while raising a number of new sociotechnical concerns \cite{bommasani2021opportunities}.

The initial set of LLM offerings from technology companies was privately designed, operated, and provisioned products. Early users of ChatGPT and competing products like Anthropic AI's Claude \cite{baiTrainingHelpfulHarmless2022} and Google's Bard \cite{liedtkeGoogleBringsIts2023} interacted with these new chatbots through gated online platforms or via APIs \cite{heikkilaFourTrendsThat}. 

While companies have made progress on advancing privately provisioned ``closed'' LLMs, various online communities have advanced efforts to build and distribute ``open'' models -- large language models that are trained on web-scale data (such as ``The Pile'' \cite{gao2020pile}) and distributed as ``raw weights'' on platforms like HuggingFace, alongside open-source code that allows users to perform LLM inferences on their own computing devices. As early as April 2022, EleutherAI's ``GPT-NeoX-20B'' was available \cite{black2022gpt}. More recently, models from Meta, Mistral AI, and others \cite{touvron2023llama,jiangMistral7B2023,jiang2024mixtral} have received attention from researchers, the media, and users \cite{goldmanMistralAIBucks2023}. The Ollama project \cite{morganJmorgancaOllama2024} further contributes by facilitating the download of various model adaptations. However, it operates primarily through a command-line interface (CLI) and is designed for individual use, limiting accessibility for non-technical users.

Although communities have been formed around these openly released LLMs on online platforms like Discord and Reddit (see e.g., GPT4All \cite{anand2023gpt4all}, h2oGPT \cite{candel2023h2ogpt}, and the r/LocalLLaMA subreddit), using these models remains a daunting task for many. The primary interface available to early adopters of local LLMs was CLIs, and using a model locally requires installation \cite{nunezLLaMAHowAccess2023, StepsGettingStarted}. This combination of growing interest and high entry barriers highlights a tension between the promise of open models and their current usability.

In theory, the proliferation of open models opens the door for alternative paradigms of LLM use. However, to fully realize this potential, platforms enabling users to run and configure interactions with open models need to be thoughtfully designed. To this end, we highlight three important design considerations for LLM hosting interfaces and toolkits. First, allowing users to run and configure open models through these platforms is already an important step in supporting openness. Openness allows people to select models based on both technical capabilities and preferences for alignment with particular values e.g., following fair training data practices \cite{fan2025can}). Second, extensibility -- which allows users to build and share their own extensions and plugins -- also requires careful design for customizing the flow of information between different models and external systems (search engines, translations, bespoke preprocessing, etc.). Third, to fully leverage both openness and extensibility, platforms must prioritize usability. For example, by lowering barriers to adoption of these platforms, simplifying installation, and making model management more intuitive, platforms can support accessibility for individual users and organizations. 

However, the reality for many users is a steep learning curve, especially without a programming or machine learning background. For instance, the release of models like LLaMa 2 generated excitement among tech-savvy communities on platforms such as Discord, Reddit, and Hacker News \cite{franzenMetaQuietlyUnveils2023}, but for many users, the process of setting up, downloading, and effectively interacting with these local LLMs remains daunting. Simplifying these processes is essential, not only to enhance user experience but also to ensure the accessibility of AI technologies.

Building on the challenges faced by everyday users, implementing open-source AI tools that can be self-hosted emerges as a critical solution. Such tools are especially valuable for people without stable internet connectivity, such as those in remote or rural areas. They also provide alternatives for users in countries where services like ChatGPT are unavailable or restricted. Moreover, open interfaces can help address privacy and data sovereignty concerns for users and organizations wary of sending queries to external data centers, a growing issue highlighted in discussions about data privacy and security (see e.g., discussion in \cite{chen2023phoenix}). Ensuring that these interfaces are open not only broadens access but also ensures that users can maintain control over their interactions and data, aligning with local needs and regulatory requirements.

The need for extensibility in AI interfaces is also becoming increasingly apparent as the field of AI grows at an exponential rate. Each user or community has distinct needs and requirements, which means interfaces must be adaptable and flexible. Implementing a robust plugin system allows for this necessary adaptability, enabling users to tailor functionalities to specific tasks or preferences. This extensibility benefits users by allowing continual adaptation to the latest AI developments, ensuring that the interface remains useful and relevant in a rapidly evolving technological landscape.

Finally, the cornerstone of broad adoption and effective use of AI interfaces lies in their user-friendliness. Simplifying complex processes is essential to make advanced AI tools accessible to a wider audience, including those with minimal technical background. Interfaces need to be intuitive, with clear guidance and streamlined processes for setting up, configuring, and using AI models. Making sophisticated AI technology easy to interact with not only enhances the overall user experience but also fosters greater engagement and democratization of cutting-edge AI capabilities, making them available and usable across diverse societal segments.

In this paper, we introduce \System{}, an open, extensible, and usable interface that allows users to interact simultaneously with many models via chat, audio, and video. Users can easily switch between local models and private model endpoints. Extensibility is a central strength of \System{}. Unlike existing systems, \System{} is designed with two distinct classes of extensibility: users can extend either the LLM capabilities or the user interface itself. These extensions can be easily shared and adopted through a community platform that is already being actively used by a large portion of the user base. Usability also distinguishes the system. Installation and configuration of basic functionalities are simplified, while heavier processing tasks are isolated in a dedicated container to ensure the main setup is lightweight. The interface also draws on the familiar design of ChatGPT to support new users while allowing for advanced functionalities and flexibility for advanced uses.

To evaluate the system, we take a multi-pronged approach. Because the system has already been deployed in the wild, our evaluation draws on real-world usage and community engagement. First, we analyze feedback from existing users across a wide variety of platforms, including feedback about the project (e.g., GitHub issues), public-facing discussion about the project (e.g., blog posts and YouTube videos organically created by users seeking to share information about the system), and publicly shared user-generated plugins. We highlight design choices that were particularly resonant with early users as well as evidence that \System achieved certain goals around openness, extensibility, and usability. Second, we report the results of a survey of active users to help understand how \System{} is meeting the design goals. 

We conclude by discussing the implications of studying a widely adopted open LLM interface. Because \System{} has been leveraged and extended by a large user community, it provides an opportunity to examine real-world interaction patterns, preferred modes of engagement with LLMs, and emerging design challenges. Our analysis also highlights how community practices such as sharing configurations and prompts shape interface use, and how local, decentralized, and open alternatives can expand opportunities for experimentation. We contribute to HCI by grounding our analysis in in the wild use and identifying lessons for designing future LLM interfaces that respond to user practices and evolving needs.

In summary, our paper makes the following contributions:
\begin{enumerate}
\item introduction of \System{}, an open-source LLM toolkit for interacting with multiple local and hosted LLMs through an open, extensible, and user-friendly interface, including a community platform for sharing and importing development resources.
\item evaluation of the \System{} toolkit through analysis of organic engagement, voluntary survey responses, and examples of how \System{} has been used in practice.
\item discussion around implications for HCI, particularly in supporting user communities, social computing, extensible human-AI interaction, and designing future LLM toolkit systems.

\end{enumerate}

\section{Related Work}
Here, we describe related work in the space of local and open-source LLMs and their use in social computing and HCI.

\subsection{Background on Local and Open-source LLMs}
Efforts to support open LLMs have been supported by grassroots contributions made via platforms like GitHub, Discord, Reddit, and HuggingFace, as well as support from private firms like Meta  \cite{franzenMetaQuietlyUnveils2023,barrMetaMadeIts}. GitHub, primarily a tool for version control and collaboration on code, facilitates user contributions through ``Pull requests" and allows community feedback via ``Issues" and ``Discussions". Discord, widely utilized for its robust communication features, including messaging and voice chat, supports vibrant communities centered on AI development. Finally, HuggingFace is a platform that supports the sharing of model weights and datasets and has achieved widespread adoption from local and open-source AI contributors.

By leveraging these online community platforms, contributors rapidly developed and shared code, datasets, and other resources that advanced local and open-source LLMs. Thus far, efforts to build open and local LLMs seem to have mirrored past successes in peer production \cite{benkler2015peer}.

A major milestone in the movement for widespread use of local and open models was the development of the open source project ``llama.cpp'' \cite{gerganovGgerganovLlamaCpp2024}, which made running LLMs practical for more users. Following this release, the ``Ollama'' project \cite{morganJmorgancaOllama2024} further expanded the reach of local models, including LLaMa and a huge array of other models, by streamlining the process of downloading and interacting with models via a CLI. 

However, prior to the advent of \System{} in 2023 \cite{openwebui}, most of these popular self-hosted interfaces were designed with cloud-hosted LLMs in mind. These platforms—while powerful and user-friendly—seldom prioritized local inferences, often requiring internet connectivity and reliance on proprietary infrastructure. Efforts to create interfaces for local LLM deployment were limited, and when they did surface, they frequently lacked the polish or extensibility seen in cloud-first systems. For instance, many early local-first solutions were constrained by usability challenges, limited support for extensions, and difficulties adapting interfaces to varied use cases. \System{} addresses these gaps by offering an interface that integrates local inference with cloud interoperability, providing a foundation that has since been adapted and extended in other open-source projects.

In this paper, we are mainly concerned with the interface for local and open models, so we use the term ``open models'' to refer broadly to any models that could be downloaded and used locally with an interface like \System.
While much of the code released by contributors is open-source, some of the licenses under which certain models have been released are not fully ``open source'' (as defined by the Open Source Initiative) \cite{OpenSourceDefinition2006}. Furthermore, there is an ongoing debate about how model weights can actually be licensed \cite{AIWeightsAre2023,contractor2022behavioral}. 

\subsection{LLMs and HCI}
Research in HCI has already begun to explore the potential of new LLM-based technologies. 

At CHI 2023, just a few months after the release of ChatGPT, there were 21 works that mentioned LLMs. By CHI 2024, this number had surged to 145, highlighting an explosive growth in the field's interest in LLMs. One notable direction of exploration has been the use of LLMs for social simulation \cite{park2023generative}, generating synthetic data for HCI \cite{hamalainen2023evaluating}, and survey \cite{xiaoTellMeYourself2020}. HCI researchers have also explored LLM interfaces, in contexts like mobile computing \cite{wang2023enabling} and prompt design \cite{zamfirescu2023johnny,arawjoChainForgeVisualToolkit2023}. All of these avenues of research stand to benefit from access to extensible LLM interfaces.

Interface design for LLMs can vary widely based on the application domain and target user group. Notably, chat-based interfaces represent one of the most intuitive and widely adopted methods for interacting with LLMs \cite{chatGPT2024,bard2024,claude2024}. These interfaces mimic human conversation, making them accessible and easy to use for a broad range of users. The versatility of chat-based UIs allows them to be adapted for various purposes, ranging from general-purpose virtual assistants to more niche applications. For instance, some users have adapted chat-based UIs to use LLMs as a virtual companion or ``AI girlfriend'' \cite{depounti2023ideal,lorenzInfluencerAIClone2023}. This application leverages the conversational capabilities of LLMs to provide users with a simulated social interaction experience. While this raises ethical and psychological considerations, it also showcases the adaptability of LLMs to cater to diverse user needs and preferences. Beyond virtual companions, chat-based UIs find their utility in numerous other applications. For instance, customer service chatbots \cite{folstadChatbotsCustomerService2019}, effective educational chatbots \cite{10.1145/3290605.3300587}, and task management assistants \cite{10.1145/3173574.3173632} all utilize the chat-based model to facilitate user interaction with the underlying AI. For these reasons, the design of \SystemSpace is motivated and mirrors these chat-based UIs. 

Prior research has also explored interfaces for LLM chaining and prototyping ML functionality through iterative prompt design. These interfaces enable users to chain multiple LLM prompts, breaking tasks into manageable steps and facilitating prompt iteration \cite{wu2022ai, wu2022promptchainer}. Systems like PromptChainer provide visual interfaces for constructing these chains \cite{wu2022promptchainer}, while Prompt Sapper extends this by offering a block-style visual programming environment for AI chains \cite{cheng2024prompt}. Similarly, PromptMaker focuses on prompt-based prototyping, enabling few-shot prompting and iterative refinement \cite{jiang2022promptmaker}. ChainForge, an open-source visual toolkit, supports simultaneous cross-LLM comparison, prompt template design, and hypothesis testing \cite{arawjo2024chainforge}. These works highlight the importance of an interactive, extensible, and user-friendly interface for working with LLMs. While they focus primarily on interfaces for prompting support—an area we also address through the Prompt Preset feature—our work broadens the interface capabilities and provide an opportunity to study how users interact with a system offering many configurable features, how they customize it to their own needs, and what this reveals about usability, interaction patterns, and emerging design challenges in HCI. 

\section{System Design}

In this section, we present the design of \System, an open-source interface and toolkit for local and private LLMs. While interfaces of this kind are increasingly common (i.e., see \cite{AnythingLLM}, \cite{lmStudio2024}, \cite{LobeChat}), our contribution lies in the system’s layered and extensible design, which enables a wide range of use cases and has supported significant community-driven involvement. 

\subsection{Current Challenges in LLM Interfaces} 
We first detail the limitations of existing UIs for private models such as ChatGPT, focusing on user privacy, limited extensibility, and restricted model selection. Then, we turn to the obstacles faced by UIs for local models, including complex setup processes, limited accessibility, and basic user interface features. These issues pose significant barriers to both research and broader user adoption.

\subsubsection{Privately Hosted UI Limitations} 
Proprietary AI interfaces like ChatGPT have several limitations. These include concerns about user privacy, lack of open-source availability leading to limited extensibility, restricted model selection, and constrained social features. First, user privacy concerns may restrict how the systems are actually used. This creates a ceiling on the contexts in which LLMs can be evaluated and audited, and could fundamentally cause certain topics and domains -- especially anything deemed sensitive -- to become understudied. Second, the ability to select from a diverse range of models would enable new HCI studies to be conducted. Such flexibility is also vital for accommodating the varied needs of different users. 

\subsubsection{Self-hosted UI Challenges} 
Self-hosted local AI Interfaces, while addressing the above concerns, suffer from their own set of issues. These include the complexity of configuration and setup processes, rendering the installation process particularly daunting for individuals with limited technical expertise, and the restriction to individual use without support for multiple accounts. Additionally, these interfaces often lack the polished user experience and advanced features common in privately developed AI interfaces. These factors collectively impede the adoption of local LLMs for wider applications and for use by everyday users.

\subsection{Design Goals for \System}
\System{} is designed to improve on both sets of challenges, and to effectively balance usability with an extensible and open approach. This involved providing a GUI that is intuitive and familiar (important for adoption \cite{yablonski2024laws}), minimizing effort to interact with the local AIs. 

Below, we provide additional details about the specific design goals for \System{}.

\subsubsection{Open}

We identified the following specific goals for \System{} that relate to taking an ``open'' approach. This entails creating a system that not only provides users with transparency into how it works but also empowers them to contribute to its development and evolution.

\paragraph{Open-source} We have made the entire source code of \System{} publicly available 
This allows users to freely examine and audit the system, its connections, and dependencies. This promotes transparency in the system and encourages collaboration among developers who can contribute to its growth and improvement. The open-source nature aligns with our goal of promoting transparency and accountability in AI development. With the entire code repository available, any user can identify potential vulnerabilities, suggest improvements, or even develop their own extensions to the system. 

\paragraph{Privacy and Data Management} HCI has highlighted privacy concerns that stem from new LLM systems \cite{zhang2024s}. \System{} is designed to prioritize user data privacy protection by implementing robust privacy controls and by being able to function entirely offline, with the option to use an API key to also query private models, unlike many interfaces for private LLMs that necessitate continuous online connectivity and logging. This design choice not only enhances user privacy but also ensures the system’s utility in environments with limited or unreliable internet access. Users have full control over their data and can choose how it is shared with third-party services or models. This includes explicit opt-in for external connections and model interactions, ensuring that users are always aware of what data is being collected, used, or shared. \System{} also stores all interaction data locally, giving users complete ownership of their data. 

\paragraph{Transparency in AI Model Selection} Unlike proprietary interfaces that often obscure the underlying AI models used, our open approach allows users to view and select from a list of available models. This enables users to make informed decisions about which model suits their needs best, promoting accountability and choice. 

\subsubsection{Extensible}

We identified three key sub-goals here: plugin support, interface customization, and integration with many models.

\paragraph{Plugin Support}

The cornerstone of our approach to achieving a highly extensible AI interface is a novel two-pronged plugin architecture that integrates front-end and back-end extensibility, while remaining script-based for full control and easy sharing through the community platform. \System{} achieves this through the prongs described below.

\textbf{Tools:} Tools within \System{} operate at the model level and allow users to make use of virtually any Python script to extend the LLM's capabilities, which supports the system's goal of being highly extensible. This not only enables real-time capabilities that require external data fetching or active interfacing with other systems, but also significantly broadens the horizons for what can be achieved with LLMs. The term ``Tools'' is also commonly used within LLM research and primarily by OpenAI \cite{openai2023chatgpt} to refer to the same functionality.

The Tools feature is designed to support a diverse range of extensibility options by allowing the integration of executable scripts that interact with external data sources and systems. In the reference implementation provided, several exemplar tools demonstrate the system's capacity to perform real-time data retrieval, execute numerical computations, and interface with dynamic state-based environments. These examples serve as foundational templates, showcasing how scripting components can be employed within the architecture to enable time-sensitive queries, perform precise mathematical operations, or fetch contextual environmental data. Such tools illustrate the extensibility of the framework, offering a flexible basis for users to develop and deploy specialized functionalities tailored to their specific research or application needs.

\textbf{Functions:} Functions operate at the application/interface level and focus on altering the functionalities of the interface itself, making them crucial for users who wish to tailor the interface according to their specific use cases.

While the term ``Functions'' is not a standard term across LLM platforms, similar concepts are often referred to broadly as ``plugins'' \cite{GitHub, LobeChat} or ``slash commands'' \cite{AnythingLLM} in other systems. However, we adopt this terminology following general programming, where a function is a self-contained piece of logic that performs a specific operation. This emphasizes that they are active, modular units of behavior within the system, distinct from just commands or plugins, which may not convey the same sense of operational logic. Functions in \System{} are organized into three distinct classes, each serving unique purposes:

\textit{Filters:} Filters serve as middleware solutions that both pre-process and post-process conversational data surrounding interactions with LLMs. Their utility spans a broad range of functions, from scrubbing sensitive information and clipping contexts to fit the operational limits of models (especially crucial for local models with restricted context windows) to monitoring dialogues to ensure adherence to usage policies. Filters can be particularly useful for supporting translation-based use cases, greatly increasing the potential pool of users who can benefit from AI systems.

\textit{Pipes:} Central to the flexibility of \System{} is the concept of Pipes, which abstracts Python functions and presents them as ``models" within the interface. This unique feature allows for an approach where Python code, irrespective of its purpose or complexity, can be seamlessly integrated and utilized just like an AI model. A Pipe encapsulates arbitrary Python functions, enabling them to interact with the system either in conjunction with LLMs or as standalone functions that bypass the need for LLMs entirely. Pipes shift the extensibility from requiring system-level updates to simply adding or modifying Python scripts, thus democratizing model integration and encouraging experimentation within the developer community.

One of the advantages of Pipes is their ability to execute independently of the traditional LLM chat interaction paradigm. This direct execution capability not only enhances the speed and efficiency of data processing tasks within \System{} but also reduces dependency on model capabilities for non-language-related tasks. Moreover, Pipes can serve as a bridge for incorporating niche models that are not natively supported under the OpenAI API spec, which is the standard API specification for most models integrated in \System{}. By wrapping these non-standard models within a Pipe, developers can bring in specialized AI capabilities (like those from newly developed or less common AI frameworks) and make them accessible to users through the same interface. This function is crucial because it allows for customization and expansion of \System{}'s capabilities without altering the core codebase. It shifts the extensibility from requiring system-level updates to simply adding or modifying Python scripts, thus democratizing model integration and encouraging experimentation within the developer community.

\textit{Actions:} Equally important are Actions, which augment the interactive elements available within the system’s UI. Actions enhance user interactivity by allowing the inclusion of customizable buttons in the response toolbar. These buttons can trigger specific tasks when clicked, offering users a direct and intuitive way to interact with the AI system beyond the standard text input. This mechanism ensures that operations requiring user consent or initiation are controlled and deliberate, maintaining user autonomy.

This plugin architecture grants users the capability to modify and enhance their interaction with LLMs in a highly personalized manner while ensuring that \System{} remains adaptable. 

\paragraph{Interface Customization} One of the key sub-goals for \System{} is to allow users to personalize their experience with features like advanced UI settings (e.g., model profile image, prompt suggestions) and custom CSS for unique themes. This adaptability ensures that \SystemSpace remains relevant and useful as user preferences evolve.

\begin{figure}[htbp]
    \centering
    \captionsetup{type=figure}
    \includegraphics[width=1\columnwidth]{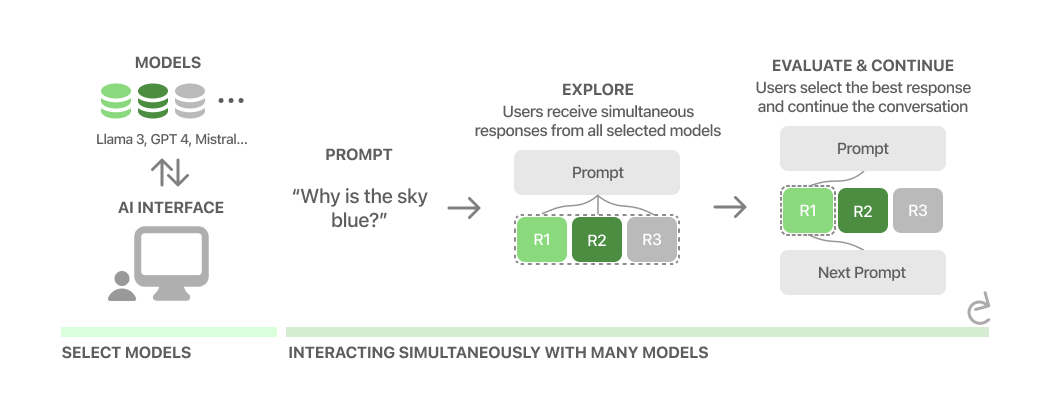}
    \captionof{figure}{Many Model Interaction: Users simultaneously interact with multiple language models, such as Llama 3.1, GPT 4, and Mistral. The user submits a query (``Why is the sky blue?''), which is processed by selected models, generating concurrent responses (R1, R2, R3). Users review and compare these responses, selecting the most appropriate one to continue the conversation. This method enables users to harness the strengths of each model, as they can select the best response for their needs and, in the process, generate preference data over models.}
\end{figure}

\paragraph{Many Model Interactions}
\SystemSpace works with both open-source and private models. This integration enables users to leverage the full spectrum of capabilities offered by these models (and new models that come out). \SystemSpace also allows for simultaneous interactions with multiple models. This broadens the potential use cases, as users can leverage the strengths of various models concurrently. For instance, a user can query a model specialized in technical knowledge, while simultaneously interacting with another model that excels in creative tasks, all within the same interface.

\subsubsection{Usability}

\paragraph{Ease of Installation and Use}
One of the primary design goals for \SystemSpace is to simplify the installation process and lower the barrier to setting up and using the system. This is achieved by automating most of the configuration process, enabling a 'plug-and-play' installation experience. The system can be installed and operational with a single command line, eliminating the need for manual configuration of environmental variables or complex setup procedures.

\begin{figure*}[htbp]
    \centering
    \includegraphics[width=0.9\linewidth]{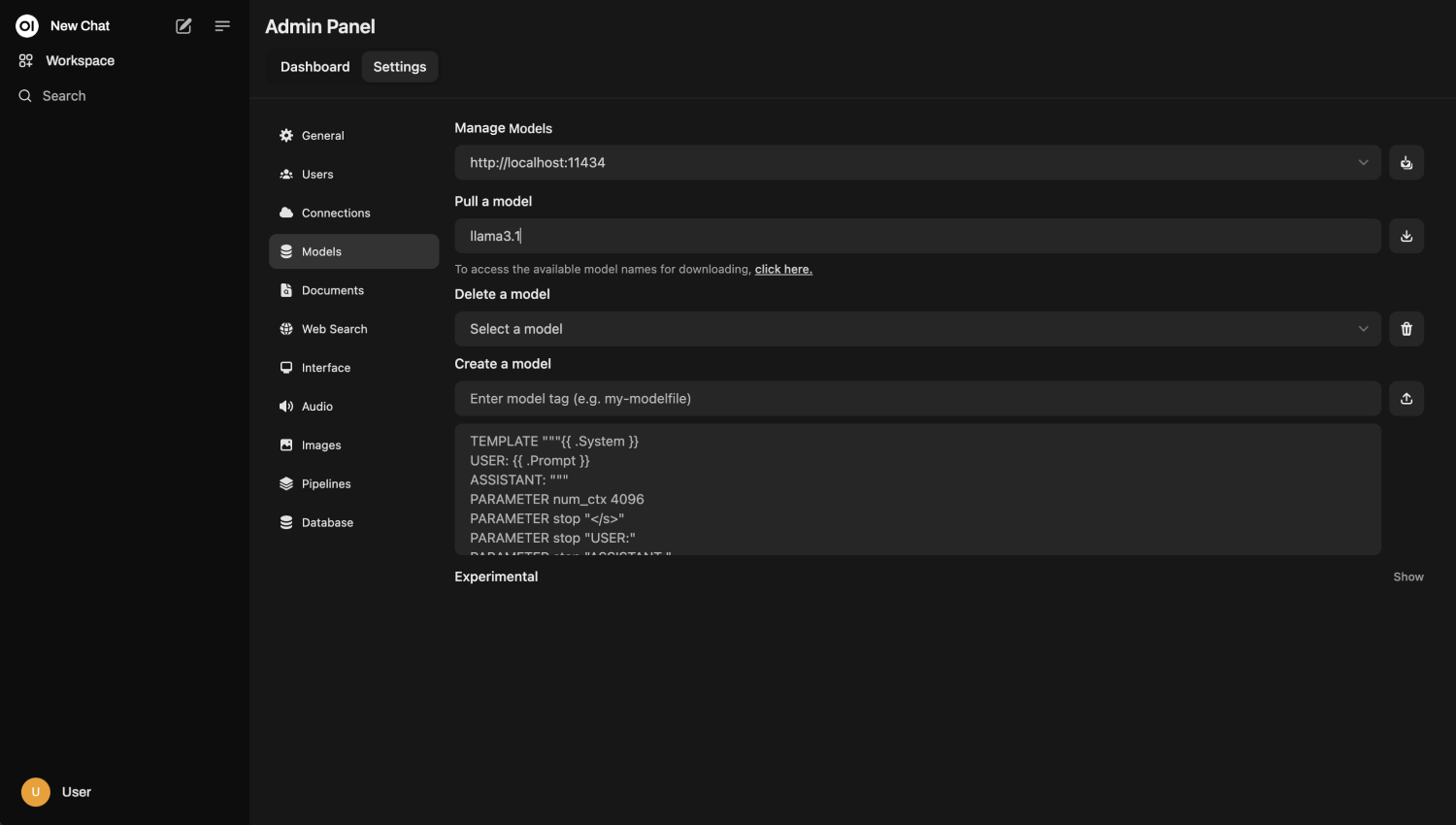} 
    \caption{\System{}'s admin settings facilitate easy AI model management through its graphical interface, enhancing accessibility by reducing dependency on command-line tools. The "Pull a model" feature enables users to download models simply by typing their names and clicking a button, with a progress bar displaying the download status. Additionally, the menu includes straightforward options for deleting models and uploading raw GGUF files, streamlining model management.}
    \label{fig:example}
\end{figure*}

Moreover, an essential aspect of \System's user-centric design involves the capability to manage AI models directly from the UI itself. This functionality streamlines the process of downloading and deploying new models, making it effortless for users to stay updated with the latest advancements in LLMs. The direct download feature negates the need for separate download and installation steps.

Finally, \System{} can also be set up within an organizational infrastructure. Installation requires only a standard server setup and an administrative account to manage user access and permissions. Employees or department members can then create their individual accounts, enabling a personalized experience while maintaining control at an organizational level. 


\subsection{The \System{} Community Platform}

The development and deployment of \SystemSpace presents an exciting new frontier in the domain of HCI, particularly in the context of social computing. Our project also incorporates a community platform, which aims to foster a collaborative environment for sharing and learning. The organization of this platform is structured around distinct resource categories: Tools, Functions, models, and prompts. Users can browse and filter these categories and seamlessly import selected resources into the \System{}, making contribution, discovery, and adoption of community content intuitive and efficient.

\subsubsection{Community Platform for Sharing and Innovation}

\begin{figure*}[htbp]
    \centering
    \includegraphics[width=1\linewidth]{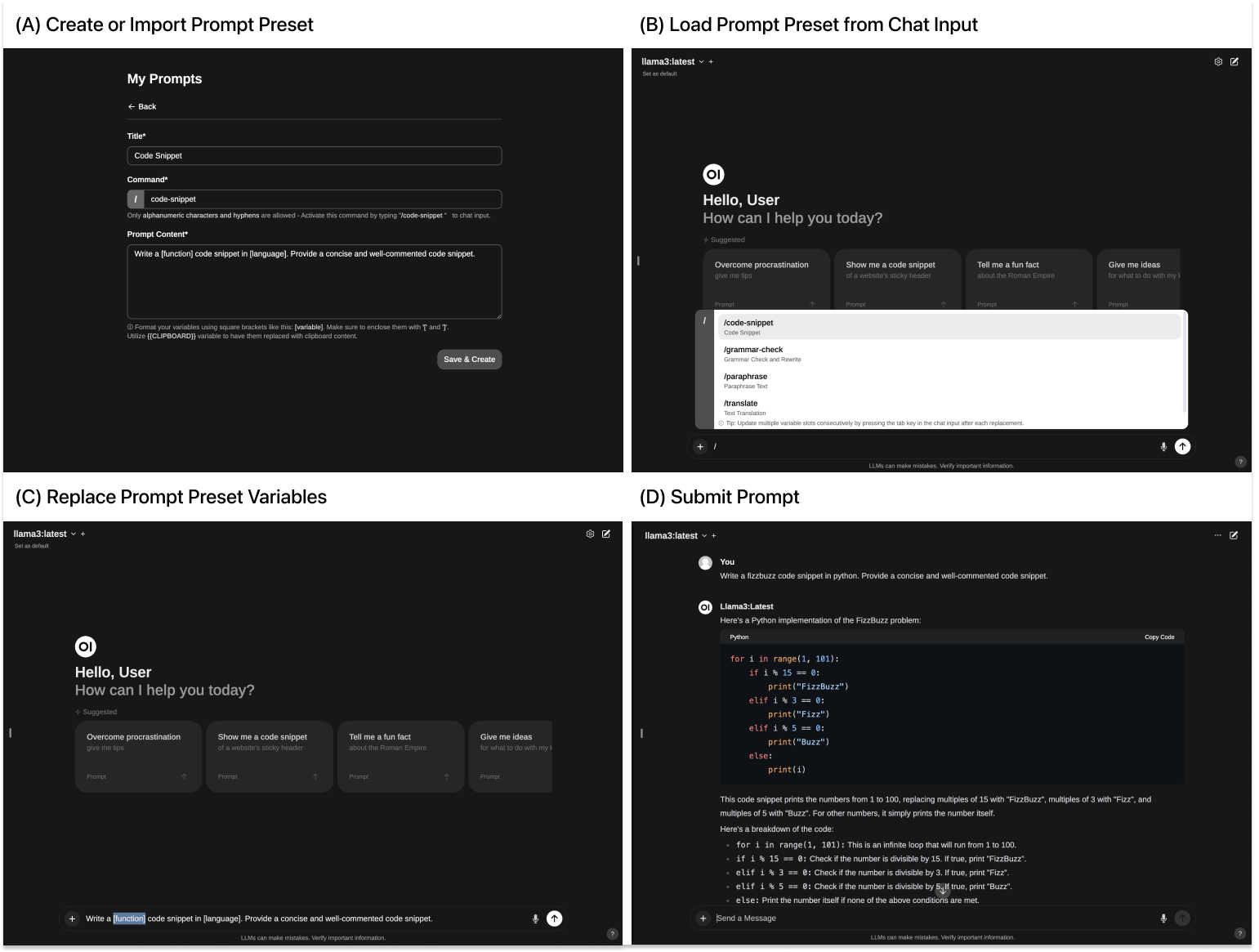} 
    \caption{\System's Prompt Preset feature, divided into four parts. (A) illustrates creating or importing Prompt Presets, where variables within square brackets are auto-selected for easy replacement using the tab key. Users also have the option to import community-shared presets. (B) demonstrates loading these presets into the chat input area using a forward slash command, enhancing user workflow. (C) highlights the efficient replacement of auto-selected variables, with the first variable selected by default and subsequent variables easily selectable with the tab key. (D) depicts submitting the customized prompt in \System, showcasing the feature's utility in streamlining repetitive tasks.}
    \label{fig:example}
\end{figure*}

The \System's community platform enables the sharing of Tools, Functions, customized prompts, and model presets with custom parameters. This design allows users to extend their models' capabilities or modify the interface without implementing these features themselves, simply by using contributions from others. It also helps non-expert users who struggle with constructing efficient prompts \cite{10.1145/3544548.3581388}, which can otherwise lead to the underutilization of LLM capabilities. The community platform mitigates this by leveraging collective intelligence, allowing users to learn from shared experiences. Consequently, this fosters an environment of continuous improvement, empowering users to refine their interactions for more meaningful LLM engagements.

\paragraph{Tools and Functions} The community platform encourages active collaboration around Tools and Functions, which were previously introduced. Users can explore extensions developed by others, experiment with them in their own workflows, and contribute improvements or variations back to the platform. This collaborative cycle accelerates the spread of innovative features and provides insights into common usage patterns and emerging needs, which reinforces the platform's role as a hub for collective creativity and knowledge sharing.

\paragraph{Custom Model Presets}
The community platform is designed to facilitate the sharing of model ``presets'' – these include custom system prompts, specific parameters like temperature, and unique UI components such as model profile images and conversation starters.

\paragraph{Prompt Presets}

Alongside model presets, users can also share custom prompt presets. These presets are particularly useful in guiding less experienced users in effectively utilizing LLMs for various tasks. By analyzing popular presets, researchers can gain insights into user preferences and typical use cases, informing future improvements in LLM design and interaction methods. This feature may significantly improve the user experience over existing privately hosted LLM UIs, like ChatGPT, which lack such sharing capabilities.

\subsubsection{Ethical Data Collection and Usage}

A notable feature of the community platform is the ability to share chat logs. This functionality is not just for showcasing model characteristics but also plays a crucial role in ethical data collection (because the system is local, data is only available from users who opt in to a particular study or data sharing pool). This aspect is particularly relevant in the current landscape, where data privacy and intellectual property rights are of paramount importance \cite{nytimeLawsuit}. 

Shared chat logs open multiple avenues for research. They can be used to study user behavior, model-user interaction dynamics, and the effectiveness of different prompts and modelfiles. This data can be instrumental in understanding how users from diverse backgrounds and with varying levels of expertise use LLMs. Such insights are invaluable in making LLMs more accessible and effective for a broader user base. In short, getting more people to use systems like \System will greatly increase the potential pool of data for academic and non-commercial research.

\section{Evaluation}
To understand how our system is meeting our design goals based on in-the-wild usage, we deployed an open-sourced \SystemSpace to the public, similar to prior work \cite{kovacs2018rotating}. We use three approaches to understand users' experience with \System{}: 1) analysis of organic public user-generated content provided by users (e.g., GitHub issues reported by users, YouTube videos, social media posts, and blogs about \System{}) through topic modeling, 2) analysis of user-contributed extensions on the community platform, and 3) a survey completed by a subset of current users.

\subsection{User-generated Content}

This component of the evaluation aims to (1) look for evidence that our three core design goals were met and (2) identify areas for improvement. Given that our objective was to create a system that is extensible and locally-focused, we look to organic engagement as a fully opt-in data source describing user experiences with the system. Similarly, as \System{} is a toolkit supported by numerous community initiatives, we first employ topic modeling on in-the-wild content to gain a broad sense of real-world use cases and behaviors. These findings are then complemented by a more in-depth analysis of specific examples presented later in the section. 

\subsubsection{Data}

We obtained two datasets for our analysis: public engagement with the project’s GitHub page and public content posted to other UGC platforms. 

\textbf{GitHub Data:} The data available from GitHub includes stars (GitHub’s `like'' feature), forks (indicating a user wants to modify and extend \System{}), user-reported issues, and discussions (forum posts on the GitHub platform). We obtained data on this engagement using GitHub’s official API in August 2024.

\textbf{Other UGC:} While public UGC has long been frequently used in HCI research \cite{blythe}, and can be a valuable potential source of data for evaluating live OSS systems, recent changes in data access and discussions about data collection \cite{freelon} – some of them stemming from LLM training data practices \cite{reddit} – complicate this process. For instance, acquiring data from Twitter and Reddit for research has become substantially more challenging.

As such, we take an approach that involves collecting snippets of UGC that have already been indexed by search engines, inspired by recent work in software engineering research \cite{wyrich} and by discussions about the ``post-API'' era \cite{poudel}.

The initial stage involved manually searching for content related to \System{} (keywords omitted for anonymity). We found content on: LinkedIn, Twitter, Reddit, Hacker News, Medium, and YouTube. This list is not meant to be exhaustive, but rather aims to cover a variety of potential perspectives. To analyze this content systematically, we collected public snippets that appear in search engine results (similar to work mentioned above \cite{wyrich,poudel} using the SerpAPI service. We collected 300 results for each platform (using queries formatted as ``site:reddit.com'') and then manually filtered this data using keywords to ensure the indexed content was directly relevant.

Poudel et al. specifically examined the viability of this ``post-API'' approach \cite{poudel}. Their results were potentially concerning regarding the use of SERP-based UGC data: results did not match random samples that were biased towards popular and positive content. However, for our purposes, we are not aiming to obtain a random sample of UGC, but rather to select for both critical and positive content and identify themes within each corpus, and so this limitation is at least partially mitigated.

After performing filtering for posts containing direct references to our system, we were left with 533 unique posts (including the title, as indexed by Google, and a snippet of content).

We treat our GitHub data and other UGC data as two distinct corpora, working from an initial observation that GitHub data seemed to primarily involve people reporting and discussing issues and pain points, whereas UGC (such as YouTube videos and blog posts) seemed to primarily involve people's tutorials, promotional content, and descriptions about how they use \System{}.

\subsubsection{Topic Modeling}
Given the varied nature of the content, topic modeling offers a scalable way to extract and compare thematic structures across different platforms consistently. 
For GitHub data, a document is the title of one issue, pull request, or discussion post. For other UGC, a document is one SERP item (page title and page snippet). While this approach necessarily misses out on specific parts of content (e.g., the main body of a blog post or the audio of a YouTube video), it has the benefit of standardizing document size across different platforms so that we can summarize topics.

Following similar methods to \cite{antoniak_blog,antoniak}, we used a Latent Dirichlet Allocation topic model. We used the Python ``tomotopy'' library with Gibbs sampling, and the data was preprocessed using regular expressions and the ``nltk'' library for removing stopwords and special characters (which we found was useful for the interpretation of our topics). We experimented with different topic numbers, using both manual evaluation of topic coherence and quantitative coherence scores. Each model was trained over 2000 iterations. After training, we used the top 10 words, along with an inspection of specific posts, to name the topics (our figures below show just the top 3 words for space).

\subsubsection{UGC Results}

First, examining the overall volume of engagement with our system, we saw that at the time of the analysis (January 2025), the GitHub repository for \System{} had 57k+ stars and 4.5k+ forks with a total of 13M+ package downloads, and 1500+ pull requests were submitted, with 320 unique contributors. Looking at our non-GitHub UGC, after using keyword filtering to find highly relevant content, we found 165 posts from LinkedIn, 147 on Twitter, 129 on YouTube, 115 on Reddit, 99 on Medium, and 24 on Hacker News. Looking specifically at our GitHub forks, several stood out as notable. Several prominent governmental organizations forked the system, suggesting the potential for locally-focused AI interfaces to support AI run by public bodies.

Additionally, our manual analysis revealed clear evidence that \System{} has already gained traction within the research community. It has been cited in studies exploring its potential for deploying LLMs in education \cite{zesch2024fernuni, eronen2024improving}, advancing applied agent design workflows \cite{ishihara2024facilitation}, and improving multilingual accessibility in special education—for example, in gloss sign language translation \cite{othman2024comparative}. A key strength of the system lies in its self-hostable and extensible architecture, which has been particularly beneficial for applications requiring local deployment to address concerns such as data privacy, regulatory compliance, or operating in resource-constrained environments. This versatility is further demonstrated in research focusing on integrating intranet and internet environments for LLMs within organizational infrastructures \cite{joadevelopment}. Collectively, these examples highlight \System{}'s ability to meet diverse user demands by fostering innovation and empowering users to design and deploy custom AI workflows tailored to their unique contexts and constraints.

Moreover, user-generated content and community experimentation provide compelling evidence of the system's flexibility for customization—both at high and low levels. At a high level, as discussed in Section 3.4, users have developed and shared plugins that extend the functionality of \System{} to interact seamlessly with external services. For example, one popular customization integrates a Google Home plugin, allowing users to control smart home devices directly via the \System{}'s web-based interface. Another user-created extension enables real-time Google Search queries from within the \System{}, effectively turning the chatbot into an interactive browsing assistant. On the lower level, some community members have gone further by adapting the system for mobile platforms, deploying it on Android devices using Termux. This not only enables local, on-device AI experiences but also demonstrates the lightweight flexibility and platform-agnostic nature of the framework. These use cases illustrate how users are actively extending \System{} beyond its original design, repurposing it for new environments and applications that span productivity, usability, and personal automation.

For the purposes of topic modeling, our GitHub data had 4926 documents (titles of issues, pull requests, and discussions), and our other UGC data had 533 documents after dropping duplicates.

\begin{figure*}[t!]
    \centering
    \includegraphics[width=.8\linewidth]{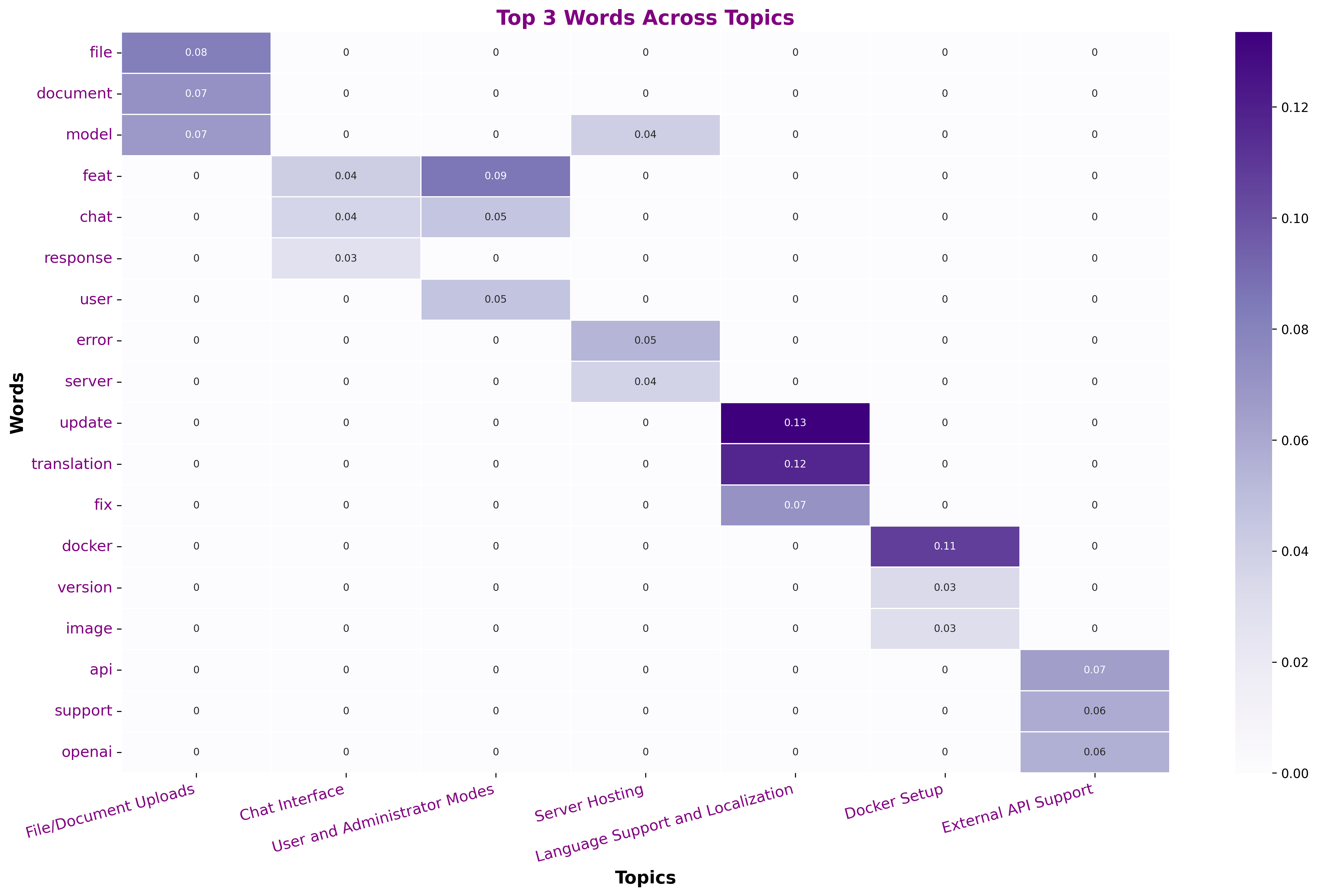} 
    \caption{A heatmap summarizing the topics in user-generated content from Github. The rows show the top three words in each topic and the columns show topics. Each cell shows how frequently a word (row) appeared in posts matching the corresponding topic (column).}
    \label{fig:github-ugc}
\end{figure*}
%

\begin{figure*}[t!]
    \centering
    \includegraphics[width=.8\linewidth]{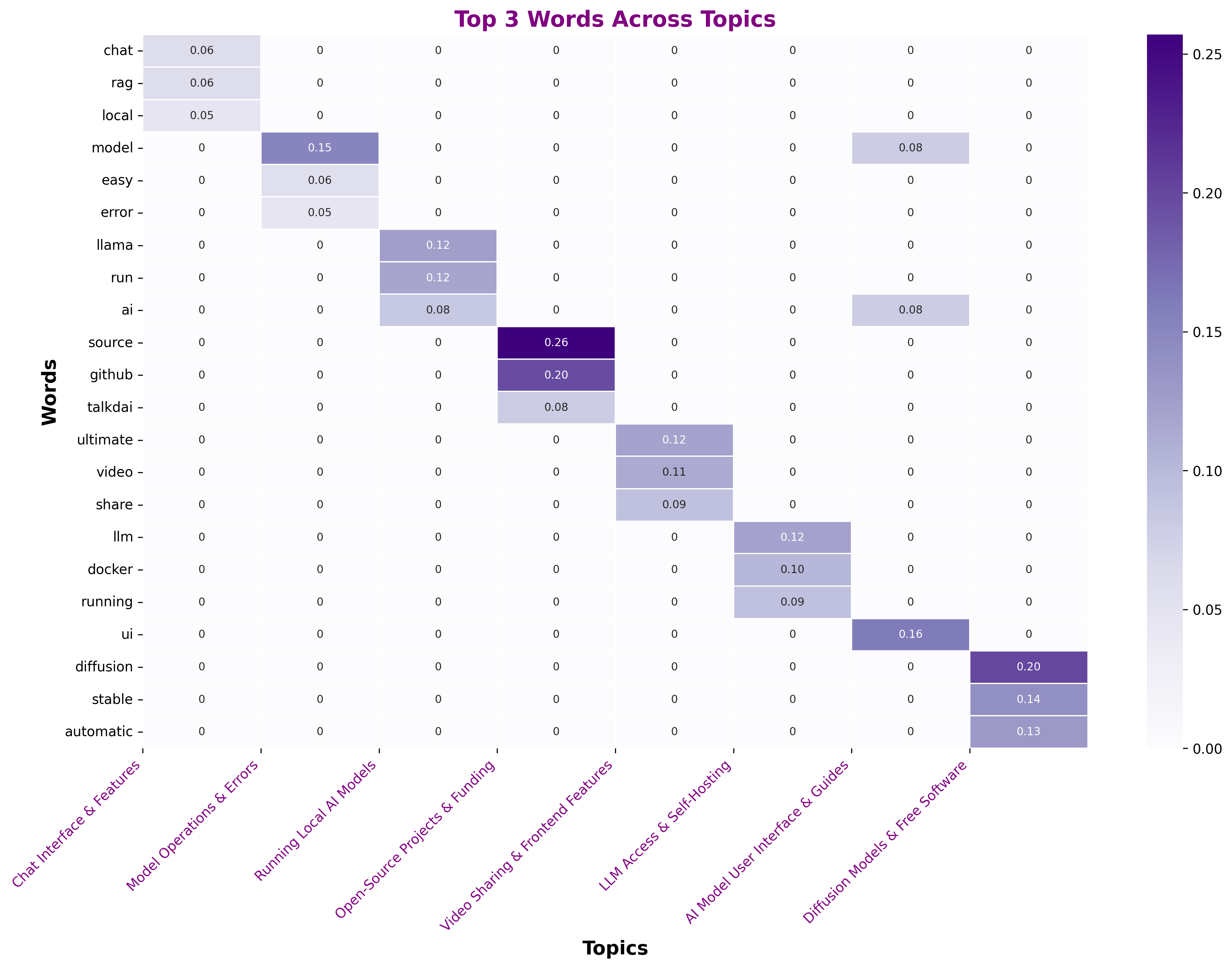} 
    \caption{A heatmap summarizing the topics in user-generated content from platforms other than GitHub (YouTube, Reddit, LinkedIn, Hackernews, Twitter, and Medium). The rows show the top three words in each topic and the columns show topics. Each cell shows how frequently a word (row) appeared in posts matching the corresponding topic (column).}
    \label{fig:ugc}
\end{figure*}

Figure \ref{fig:github-ugc} summarizes topics we identified in our GitHub data, while Figure \ref{fig:ugc} covers the UGC data. Each figure shows the top three words from each topic and the labels we assigned each topic based on their top words and key examples.

Looking at these topics in our GitHub data, a key finding was the emphasis on user desire for features that work with local data (``File / Document Uploads''). This resonated with our manual examination of feedback data, in particular, a frequent request by users for ``Retrieval Augmented Generation'' (RAG), a specific technique in the generative AI space for working with specific documents. In this case, the presence of a topic provided very straightforward feedback: to satisfy user demands, open-source projects have to keep up with new capabilities and features. 

Other topics appearing in GitHub engagement included discussion about chat interfaces, server hosting, and discussion of specific issues with Docker and external APIs (e.g., using \System{} with the OpenAI API). These suggested that users were using our system in a variety of contexts.

In our models summarizing other UGC data, we also see discussion by users of features such as RAG (leftmost topic), discussion on the local emphasis of our system (third from left), as well as mentions of specific models (stable diffusion, rightmost topic). It is also notable that in the content they posted, people directly discussed open-source funding, a key consideration for the sustainability of similar projects.

Looking manually through our UGC data, we observed broadly four different types of content: tutorial, demonstration (including direct examples of extensibility), testimonial, and integration. In terms of standout examples of UGC, we identified YouTube videos that demonstrated particularly in-depth engagement with \System{} (in terms of the effort to create the video, and the engagement with the video itself). Tutorials on YouTube covered installation, how to use multiple AI models, how to work with prompts, how to implement RAG, voice input, and more. This observation suggested that emphasizing the open-source aspect of \System{} was, as one might expect, helpful in allowing users to participate in making the system more usable. 

To summarize, analyzing organically generated content from users suggests that \System{} was able to meet the demand for a local-focused, privacy-friendly interface for using AI models, but that technical issues (e.g., installation) are likely still a barrier to more widespread use. As expected, content posted to GitHub focused on issues whereas content posted to other platforms focused more on promotion, tutorials, and positive feedback. We identified direct areas for improving our system's usability (address issues with installation, hosting, Docker, external API support) and feature set (continue adding features that support local-first interaction with documents). These findings offer valuable guidance for HCI researchers developing similar toolkit platforms and highlight the practical aspects that directly affect adoption and user experience.


\subsection{User-Generated Plugins}

In addition to engaging with the project on GitHub and posting in-the-wild content on platforms like YouTube and Reddit, \System users also utilized the extensibility of the system by building and openly sharing user-generated plugins: Filters, Pipes, and Actions. The community-driven ecosystem surrounding \System{} has generated a considerable volume of these plugins (this volume is a useful indicator that the goals of openness, extensibility, and usability resonated with users). Currently, the user community has contributed 541 unique functions and 276 unique tools to the community platform, with additional plugins available through platforms like GitHub. In this section, we offer a comprehensive overview of notable examples of these user-generated plugins, demonstrating the utility, innovation, and breadth of applications made possible by \System's extensibility. 

In order to understand how users utilized the extensibility of \SystemSpace, we also applied topic modeling to the corpus of user-contributed plugins. Our preprocessing and modeling approach is consistent with that described in Section 4.1, but we extend the stopword list to include terms frequently appearing in plugin descriptions such as ``tool'', ``function'', ``feature'', and ``application''. For each identified topic, we select representative extensions based on their probability scores. Our focus in this section is not on reporting the exact distribution of topics, but instead on systematically answering, ``What did users do with the extensibility offered by \System{}''. We include raw model outputs in an Appendix.

To better contextualize community contributions, we also report each user-generated extension's number of downloads recorded from the community platform and rounded to the nearest hundred or thousand. 

This analysis offers several advantages in this context. It provides a reproducible method to identify meaningful groups of functionality, highlights patterns in use cases, and reduces manual bias in categorization.

\subsubsection{Tools}

Tools are designed as powerful extensibility mechanisms, allowing users to integrate commands that extend beyond traditional LLM capabilities. By executing external Python scripts or interfacing directly with APIs and external services, these tools significantly augment the interactive and computational possibilities of local and privately hosted LLMs. 

We performed topic modeling on the top 100 tools on the community platform by download count to find themes and examples that are representative of the community's most influential extensions.

Some notable plugin tools contributed by the user community include:

\begin{itemize}
\item \textbf{Enhanced Web Scrape [16K Downloads]} — A web scraping tool that extracts text content from web pages, supporting user customization and improved filtering.  
\item \textbf{ComfyUI Image Prompt [552+ Downloads]} — A tool that converts images into prompts for image generation workflows, enabling enhanced image-based prompting.  
\item \textbf{WolframAlpha LLM API [2.6K Downloads]} — This tool uses the WolframAlpha LLM API to access knowledge and retrieve information.  
\item \textbf{Run Code [6.2K Downloads]} — Executes Python or Bash code securely within a sandboxed environment.  
\item \textbf{SQL Server Access [2.9K Downloads]} — Provides access to SQL databases, allowing users to query, retrieve, and explore database content.  
\item \textbf{Stock Reporter [7.6K Downloads]} — Gathers stock market data and generates comprehensive reports.  
\item \textbf{Deep Research (Browser UI) [1.5K Downloads]} — Performs real-time research and data retrieval from online sources.  
\item \textbf{Home Assistant Light Control [642 Downloads]} — Enables control and management of smart lights through the platform.  
\end{itemize}

These tools strongly support the stated goals of \textit{openness} by interfacing with external APIs and sources, while through \textit{extensibility}, they provide essential new functionalities that extend model capabilities beyond typical language-based tasks.

\subsubsection{Functions}

Functions provide modular ways to enhance or customize the behavior of the \System{} system itself. Recall, these are composed of Filters, Pipes, and Actions, each of which serves unique and clearly delineated purposes. 

For the topic modeling, our analysis was limited to the top 30 functions for each class due to the smaller overall set compared with the tools. Only three items were selected per class according to their highest probability scores for brevity.

\paragraph{Filters}

Filters act as middleware, refining I/O interactions on their way into or out of the LLM interface. Exceptional community-contributed examples, reflecting diverse objectives, include:

\begin{itemize}
    \item \textbf{GPT Usage Tracker [1K Downloads]} — Tracks usage and costs for GPT models, allowing users to monitor and log model interactions.  
    \item \textbf{Google Translate [4.1K Downloads]} — Provides automatic translation between a user's preferred language and the language used by the LLM.  
    \item \textbf{AutoTool Filter (User Setting) [145 Downloads]} — Preprocesses user queries to identify and select relevant tools automatically.  
\end{itemize}

Filters directly support our design objectives of \textit{openness} and privacy by controlling and managing conversational data flows. Specifically, community-generated Filters addressing anonymization and personally identifiable information (PII) redaction align closely with our stated privacy and data protection goals.

\paragraph{Pipes}

Pipes allow arbitrary Python functionalities to behave analogously to native models in \System{}. This greatly simplifies and generalizes how users interact with different models or integrated services. Some user-created Pipes notably include:

\begin{itemize}
    \item \textbf{Anthropic Claude Model Access [327 Downloads]} — Provides API access to the Anthropic Claude models, supporting the latest API features such as prompt caching, document input, and multimodal capabilities.  
    \item \textbf{DeepSeek R1 Think Chain [695 Downloads]} — Shows the reasoning chain of the DeepSeek R1 model, helping users follow and understand model decision processes.  
    \item \textbf{DeepSeek V3 R1 Gemini Vision [558 Downloads]} — Integrates DeepSeek with Gemini Vision for enhanced image analysis and processing, combining advanced vision capabilities with precise reasoning.  
\end{itemize}

The Pipes promotes \textit{extensibility}, opening \System{} to myriad AI services, modalities (text, vision, images), and cross-model benchmarking.

\paragraph{Actions}

Actions are UI-driven functionalities that allow direct, explicit invocation by users, complementing more traditional text-only inputs. Key user-generated Actions include:

\begin{itemize}
    \item \textbf{Mixture of Agents [9.2K Downloads]} — Enables combining the strengths of multiple models in a layered, iterative workflow to improve response quality and decision-making.  
    \item \textbf{Visualize Data [13K Downloads]} — Generates charts from conversation data, allowing users to quickly view and interpret information within the chat.  
    \item \textbf{Add to Memories [9.1K Downloads]} — Saves assistant messages to the user’s memory, supporting persistent context for ongoing interactions.  
\end{itemize}

Actions directly tie into \textit{usability}, significantly reducing user friction in performing additional tasks within \System{}.

Ultimately, examining community-contributed plugins complements the prior UGC analysis by specifically exploring the range of use cases the toolkit can support and highlighting extensions most representative of key functional areas of the platform. We discuss further implications in Section 5.

\subsection{User Survey}
In addition to analyzing the organic engagement with \System, we created and posted a voluntary non-paid survey for current users to complete to better understand how the \SystemSpace is currently being used and areas of improvement. The survey was 
approved by our institution's ethics review board.

\subsubsection{Methods}
To gather user feedback, we created an online survey and announced it on the dedicated Discord channel, \System{} subreddit, and GitHub Discussions section for the diverse user community within the \System. The survey was designed to solicit input from users spanning a range of use cases and expertise levels, including developers, end users, and researchers actively engaging with the system. The survey aimed to capture how user interact with \SystemSpace, including their use of extensibility features, and overall system functionality. 

The survey was structured to collect feedback across five areas: (1) user background and experience with LLMs, (2) how participants use \SystemSpace and their motivations for choosing it, (3) experiences with extensibility features such as tools and functions, (4) perceptions of usability, and openness, and (5) overall experience. We included these questions to capture general user feedback and also explicitly assess whether the system meets its three primary design goals through a combination of quantitative and qualitative questions.

The survey included a combination of question types. Likert scale questions measured familiarity, ease of use, and perceptions of transparency/openness. Multiple choice questions captured users experiences with \System{}, their uses, and reasons for choosing it. Free-text responses allowed users to provide further details on ease of use, projects that used the extensibility features, impact of openness on their experiences, as well as likes and dislikes. All responses were manually reviewed to identify overall themes for each question. Representative examples from identified themes were reported. For multiple-choice questions, responses outside the predefined options were reported, and for free-text questions, unique or more detailed answers were presented.

Within two weeks starting in August of 2025, we received a total of 20 survey responses from active users from 10 countries: Argentina, Australia, Austria, Brazil, Canada, Germany, Norway, Singapore, the United Kingdom, United States of America. We did not ask any other demographic questions to help preserve privacy. On average, the survey took 5-10 minutes to complete. 

\subsubsection{Results}

\begin{figure*}[htbp]
    \centering
    \includegraphics[width=1\linewidth]{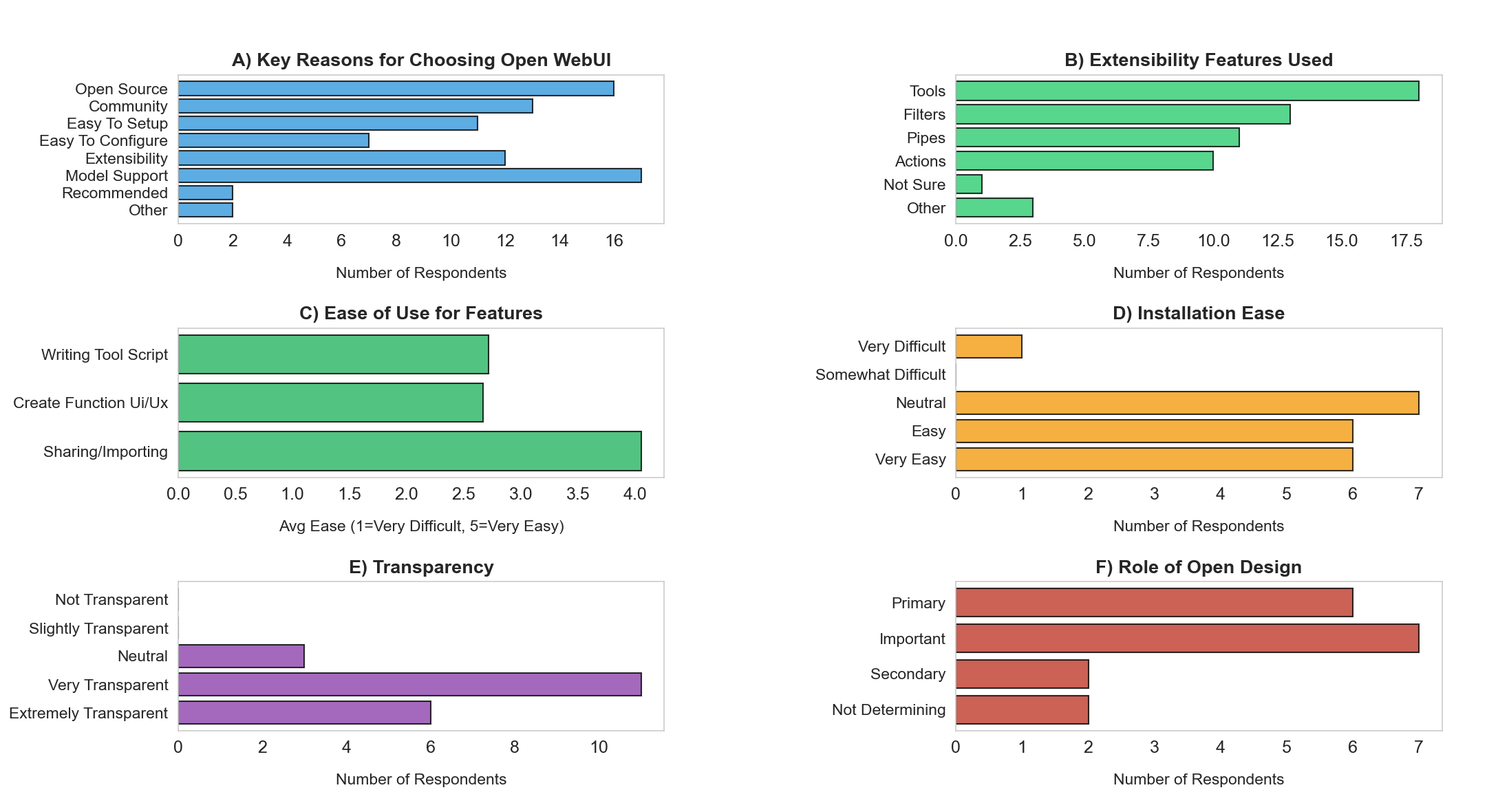} 
    \caption{Survey results showing reasons for choosing \System{}, usability of extensibility features, ease of installation, perceptions of transparency, and the role of open design.}
    \label{fig:survey-plots}
\end{figure*}

\begin{figure*}[htbp]
    \centering
    \includegraphics[width=1\linewidth]{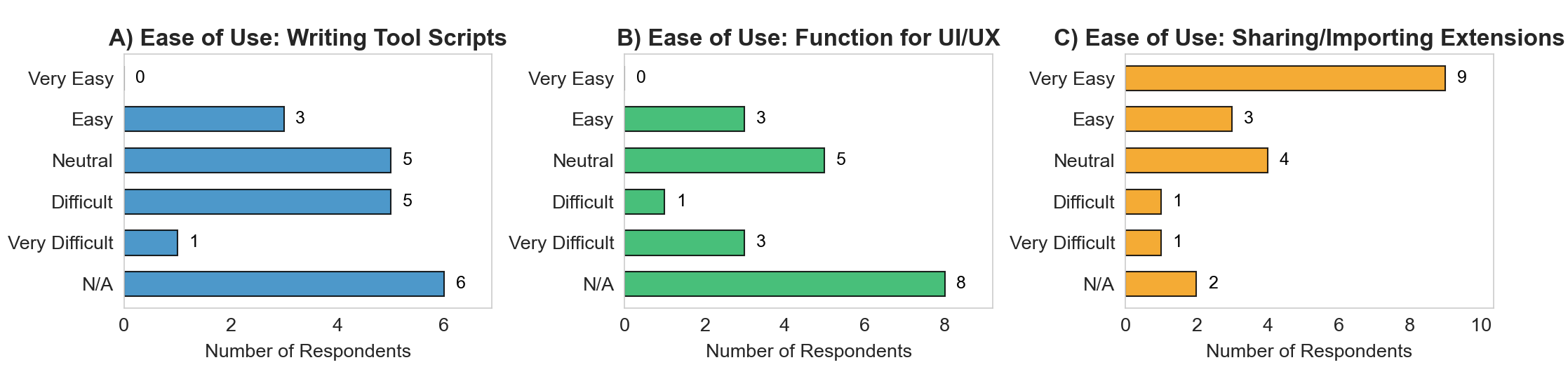} 
    \caption{Survey results showing the difficulty of using each extensibility feature.}
    \label{fig:survey-plots-diff}
\end{figure*} 

Here, we provide a summary of participants' responses to each question.

\textbf{User Background and LLM Experience}
These questions collected basic information about participants, including their familiarity with LLMs, and prior experience using or configuring them. The goal was to contextualize responses by understanding the diversity and expertise of the user base.

Respondents reported a high level of familiarity with LLMs, with an average rating of 4.4 on a 5-point Likert scale. Regarding prior experience, 18 respondents had interacted with LLMs via a user interface, 18 had configured or set up LLMs on their own systems, 9 had created extensions, and 15 had integrated LLMs into other systems or workflows. Most participants had hands-on experience across multiple aspects of LLM use, providing a solid foundation for evaluating \SystemSpace.

\textbf{Usage Patterns and Motivations}
Participants were asked about how they use \System{} and why they chose it. This included questions about the types of tasks they performed and their key motivations in order to identify common usage patterns and features of the platform users find most valuable.

For usage of \System{}, respondents reported a variety of activities: 15 used it for code generation or debugging, 14 for writing or creative work, 17 for research, 12 for personal assistant or automation tasks, 13 for teaching or learning about LLMs, 12 for integrating with external tools or APIs, 12 for recreational uses, and 1 respondent indicated a custom use — ``summary and preliminary analysis of different data sources.''

Regarding the key reasons for choosing \System{}, respondents highlighted multiple motivations (see Fig. \ref{fig:survey-plots}A): 16 chose it because it is open-source, transparent, or designed to protect user privacy; 13 valued the active community and contributor-friendly design; 11 cited ease of installation and setup; 7 noted ease of configuration and customization; 12 appreciated the ability to extend with plugins; 17 valued support for multiple LLM models and providers; 2 selected it because it was recommended by others; and 2 provided custom reasons, including: ``It’s a great piece of software. Before learning about \System{}, I tried coding my own UI. Then immediately stopped once I figured out how to deploy WebUI + Ollama in my homelab,'' and ``Integrated user role system and permissions, Knowledge base system.'' These responses demonstrate that users are motivated by a combination of openness, extensibility, and multi-model support, which aligns with the system’s design goals.

\textbf{Experiences with Extensibility Features}
Tools were the most widely used extensibility feature, reported by 18 participants, followed by Filters with 13, Pipes with 11, and Actions with 10 (see Fig. \ref{fig:survey-plots}B). 

We also asked about projects directly using the extensibility feature and found that 15 of 20 responses described projects or experiments, while 5 participants stated they had not created anything. Four participants mentioned working with Model Context Protocol (MCP) tools or planning MCP integrations, while one built automation features such as deep research, email, and Teams messaging. One person described a central knowledge hub for a marketing team, another added extra context for models with small memory windows, and one created a news feed scraper and summarizer. Individual contributions also included a Document360/Google Drive sync for a knowledge base, LaTeX whitepaper generation for retrieval, code workflows, and basic prompt modifications, such as adding time stamps or adjusting instructions. Three participants reported smaller or unfinished work, including an AssemblyAI transcription attempt, custom filters, configuration of TTs, and use of ChatGPT’s API, while two simply shared GitHub links to their code.

Participants also rated their experience with different extensibility features. For a visualization of the distributions, see Fig. \ref{fig:survey-plots-diff}. For ease of use, writing tool scripts were rated with a weighted average of 2.7 and a standard deviation of 1.0. This shows that most participants found this task to be of moderate difficulty rather than straightforward. 
Creating functions for UI/UX changes was rated with a weighted average ease of use of 2.7 and a standard deviation of 1.3.
Here, similar to the Tools results, participants were split, with several finding it difficult and only a few reporting ease of use. Sharing or importing extensions from the community platform had a weighted average ease of use of 4.0 with a standard deviation of 1.3. This indicates that the majority found community extension sharing relatively accessible compared to other extensibility tasks. 
Our findings show that community-driven sharing mechanisms are comparatively more accessible and easy to use, in contrast to the difficulties users reported when developing or customizing their own extensions.

\textbf{User Perception of Usability and Openness}
This final set of questions elicited user feedback on usability, transparency, openness, and comparisons to other LLM interfaces, along with general impressions and suggestions for improvement. These questions were included to assess whether \System meets its design goals of openness and usability.

For ease of installation, the weighted average was 3.8, with a standard deviation of 1.2.
(see Fig. \ref{fig:survey-plots}D). This shows that most participants (12) found installation straightforward, though a smaller group experienced challenges.

Open-ended responses about the setup process reported a mix of straightforward and difficult experiences. Ten participants generally found the initial installation simple, especially when using Docker Compose or the one-command setup, and three noted that adding models was easy. However, several highlighted difficulties: 2 users reported challenges with networking and proxy configuration, poor or outdated documentation (2 participants), and challenges with advanced features such as RAG, MCP tools, and API integrations (5 participants). Two participants also mentioned struggles with permissions management and consistency in handling uploaded files. These responses indicate that while \System{} lowers the barrier for getting started, more complex workflows remain difficult to configure.

When asked about transparency, no participants rated the system as ``not transparent'', instead, the average transparency rating was 4.1 with a standard deviation of 0.6.
This suggests that most participants felt the platform provided strong transparency into model use, API controls, and data handling, as shown in Fig. \ref{fig:survey-plots}E.

On the role of open design, participants overwhelmingly emphasized its importance. Many described it as a ``primary driver,'' ``critical,'' or ``mandatory'' for adoption (6 responses), while others highlighted specific benefits such as owning their stack, model agnosticism, or privacy guarantees. Two participants also noted that the open-source nature and ``beautiful UI'' differentiated \System{} from alternatives and made it their preferred choice. Only four participants saw it as secondary or ``not a determining factor.'' For full visualization, refer to Fig. \ref{fig:survey-plots}F.

\textbf{Overall Experience}

Participants consistently praised the open-source and extensible nature of \System{}, with 8 respondents explicitly mentioning extensibility, openness, modularity, or customizability as a strength. For example, one participant stated: ``Degree of customization, aggregate multiple LLM providers. Extremely easy to switch to another model/company when a new SoTA is released (which is very often in the current stage of LLMs).'' Another emphasized flexibility, stating: ``It can do whatever I want it to.'' The user experience and interface was another common theme, highlighted by 7 participants, who described the UI as ``good,'' ``decent,'' or ``nice.''. Four participants emphasized the ability to integrate or switch between providers, while community involvement was valued by 3 respondents. Furthermore, control and ownership of data were noted by 2 participants. Overall, the most consistent positives were its flexibility, customization, and strong UI, coupled with the value of being an open-source project supported by an engaged community.

Dislikes were more varied, but several themes emerged. Documentation and learning curve issues were mentioned by 5 participants, who found features hard to understand, poorly documented, or confusing in implementation (e.g., RAG, system prompts, tool use). UI customization limitations and performance issues were highlighted by 4 participants, ranging from a lack of theme options to frontend bugs. Another 3 participants pointed to installation or configuration difficulties, while 2 expressed concerns about licensing restrictions and enterprise integration. Two participants also mentioned limited features, such as TTS or markdown support. Lastly, 2 participants reported having no significant dislikes or mentioned some dislike for occasional bugs. In summary, the strongest areas of concern were documentation gaps, UI performance, customization limitations, and installation challenges.

Responses comparing \System{} to other platforms were strongly favorable. 13 participants explicitly stated that \System{} is better or the best among platforms they have used, often citing extensibility, customization, and feature richness. For example, one participant stated: ``It has no direct competitor, all other options are behind in features and extensibility.'' Another highlighted flexibility compared to proprietary platforms: ``Much better than using ChatGPT’s interface, I like that I can choose the tools to use and the connections to make, and how I can export conversations and choose my database.''
Four participants contrasted \System{} positively against specific alternatives, such as LibreChat and AnythingLLM, noting that while these competitors sometimes offered a more polished interface, they lacked \System{}'s breadth of features. As one explained: ``Tested only AnythingLLM... while their presentation and UI look more polished and complete, it lacks many features \System{} has. The user groups and permissions was really the feature that moved us to try \System{}''. Three participants reported limited use of other platforms, stating they had little basis for comparison, and 1 participant raised licensing as a drawback despite otherwise viewing \System{} as favorable. Overall, the majority of participants saw \System{} as preferable to mainstream interfaces and competing open-source projects, particularly because of its customization, integration options, and active development.

\section{Discussion}
Above, we discussed the design of \System, provided a list of social computing applications that \SystemSpace enables, and reported on an early evaluation of our approach, with a focus on how well the system achieved three design goals: simultaneous openness, extensibility, and usability. Here, we discuss several potential consequences of future efforts aimed at making local LLM interfaces easier to use, including benefits for users across various geographic contexts, communities with diverse values, and the benefits of decentralization. We highlight fruitful directions for future work along each of these lines.

Our early evaluation results suggested that the initial design of \SystemSpace made significant progress towards meeting its design goals. Early users also reported high levels of experience with LLMs (and self-selected to engage with GitHub, Discord, etc.). Therefore, we encourage future work to explore further how interfaces like \SystemSpace may better support the needs of an increasingly broad user population.

\subsection{Design Implications for LLM Toolkits}
Overall, this work contributes to HCI by revealing how users engage with and extend an open-source LLM toolkit platform. To our knowledge, there has been little HCI work documenting user experiences in an open-source LLM toolkit at scale, making this study an early step toward understanding and highlighting how these interfaces can be designed to support extensible human-AI interaction.

We find that participants consistently value the openness of the platform, the multi-model support, and the level of customization provided. These features were seen as key reasons for choosing \System{} over proprietary systems. However, participants also mentioned several challenges related to customization, such as creating new plugin scripts from scratch. In contrast, participants found using community contributions much easier. This points to the important role of the community platform in extensible AI systems for lowering entry barriers and enabling easier experimentation, where the value of the platform is multiplied by users' ability to share and build upon each other's work. Furthermore, HCI research should continue to explore ways to simplify the process of using advanced functionalities, such as creating plugins in these toolkits.

We also show the importance of careful interface design of these platforms. Participants valued the clean and intuitive UI, but noted challenges with documentation, configuration, and performance. This reinforces the need to balance extensibility with usability to ensure users can navigate, understand, and use these customization features with appropriate support.

We see several opportunities for future work. First, we encourage HCI researchers and LLM platform developers to engage in more systematic studies of user contributions, experiences, and practices in open-source settings to better understand how to advance these platforms. Second, future research can explore how extensible architectures may be designed to better support both expert and novice users, either through improved methods for onboarding, documentation, or new forms of collaborative extension development. Lastly, further work is needed to understand the social dynamics of \System{}, including how users collectively improve the platform, assist one another, and how the interface itself supports this active, collaborative community -- features largely absent in proprietary platforms. We hope our work serves as a stepping stone to guide the design of future LLM toolkit systems.

\subsection{Additional benefits of openness and extensibility for ``local'' models}

\textbf{Geographic inclusion:} We observed that survey participants hailed from a variety of countries across the world. This suggests that open interfaces can support users in locations that limit access to private LLM offerings. Two major reasons someone might be unable to use an LLM service like ChatGPT or Bard might be local regulations, cost, or lack of stable Internet access (e.g., people living in rural areas or developing countries). In the future, attempts to deploy LLMs in extreme environments (e.g., space) might also benefit. Similarly, organizations stand to benefit from running local LLMs with no outbound data, especially given the availability of a local model whose performance is on par with ``GPT-3.5." \cite{zheng2023judging}.

\textbf{``Many models'' approach:} Furthermore, an open and extensible approach means that users can benefit from many models at once, each potentially containing domain-specific advantages or tuned to a particular set of human values. Individuals could download specialist LLMs to their local devices, enabling access to expert knowledge in various fields, including medicine, engineering, and law. The medical field, in particular, may benefit significantly from this development \cite{nori2023capabilities, 10.1145/3458754,tu2024conversational}. Local LLMs may serve as an initial point of contact, offering preliminary diagnoses or health advice, thereby reducing the strain on healthcare systems and freeing up specialists' time for more critical cases. Of course, mitigating risk will require attention to the challenges of using LLMs in high-stakes scenarios \cite{mesko2023imperative}.

\textbf{Pluralism through online community governance}: The platforms where users are discussing and sharing \System and other local AI tools (e.g., GitHub, Reddit, Discord, HuggingFace, the \System social platform) are of great interest in social computing, especially scholarship on online governance. These platforms require moderation practices \cite{jiang2019moderation} (sometimes using bots \cite{kieneWhoUsesBots2020,kieneTechnologicalFramesUser2019}), and sometimes use governance tools (e.g., \cite{zhangPolicyKitBuildingGovernance2020}) to allow communities to employ voting, juries, and more. Reddit \cite{baumgartnerPushshiftRedditDataset2020a} and GitHub \cite{maldeniyaHerdingDelugeGood2020} are especially well studied in social computing, but further understanding Discord and any new social platforms will be critical for understanding the future of AI systems.

In a world of many models -- collaboratively built and governed through interactions using online platforms -- the design of these platforms could be shaped to support a pluralistic approach to \textit{values}. There remain many open research questions about how models themselves might achieve values pluralism \cite{sorensen2023value}, but it seems likely that online communities can foster broader public engagement with AI governance.

\subsection{Future Work and Limitations}

Our results suggest that building more open, extensible interfaces for LLMs can enable a number of new directions in HCI research. One particularly exciting direction of research might be applying systems like \System{} to the evaluation of AI. The rapid advancement of LLMs has created an ``evaluation crisis'', underscoring the need for effective, ethical, and socially responsible evaluation and auditing methods \cite{HEAL}. Extensibility can help users to participate in the evaluation of AI, as a system that is flexible enough to accommodate diversity of use cases and user groups, especially in local settings, can actually collect data needed to evaluate AI that otherwise might be expensive or impossible to acquire via a ``centralized'' private approach.  In the case of \System, this involves creating frameworks that allow users to assess the model's performance based on parameters that are significant to their specific context.

Above, we noted that the choice to use local LLMs, for individual users or for organizations, will involve navigating trade-offs. Here, we reiterate some of the expected limitations of the local LLM  approach and the current implementation of \System.

First, while some users may see cost savings, ultimately, all LLM users must either pay for API access or pay for compute resources (and associated costs such as energy, maintenance, etc.). Researchers have begun to highlight concerns with the energy use of AI \cite{schwartz2020green}; it is likely that a shift toward local models may decentralize decisions around energy use and compute costs.

Second, we note several system-specific limitations. Survey respondents identified challenges related to installation, server deployment, and hardware constraints. Although \System{} was commended for simplifying these processes, participants noted that familiarizing oneself with the system for the first time, and configuring or creating custom extensions, can be challenging. This is especially true as \System{} and many other LLM platforms implement extensions through scripts, which are straightforward to import but require some technical expertise to create from scratch. 

These findings highlight several directions for future work. First, simplifying system installation and the creation of extensions to better support non-technical users or those without development experience would broaden the accessibility and usability of \System{}. Although many existing extensions cover the majority of typical use cases, providing novice users the ability to develop their own extensions easily for highly custom or unique workflows remains important. Second, conducting a larger-scale survey would provide a more comprehensive understanding of user experiences. 

\section{Conclusion}
In this paper, we introduced \System, an open-source LLM toolkit that streamlines the use of local large language models (LLMs) by providing a user-friendly interface for downloading, installing, and managing various models. \System{} supports both open-source and proprietary models and provides a collaborative ecosystem in which users can share and import community-contributed extensions and presets. Our evaluations, based on organic user engagement, survey responses, and curated examples of toolkit use in the wild, indicate that \System{} is making significant progress toward its design goals of openness, extensibility, and usability. We highlight directions for future work, including simplifying installation and lowering the barrier to creating extensions. Overall, our findings highlight the potential of LLM toolkit platforms and encourage both research and platform developers to contribute to HCI studies that inform the design of future local LLM interfaces.

\section{Acknowledgments}

We thank the global open-source community whose creativity and collaboration have been central to \System{}'s development. The efforts of independent maintainers and contributors worldwide, working across borders and areas of expertise, exemplify the potential of open-source collaboration. We also recognize and appreciate the efforts of Open WebUI, Inc. for its role in supporting the broader ecosystem and making \System{} publicly available.\footnote{Use of the software is governed by the license terms in the official project repository.}

This paper is an independent academic study of \System{} in the context of AI and HCI research. The authors’ contributions here are limited to analysis and description. Development and maintenance of \System{} are carried out independently by the community and Open WebUI, Inc. Any personal contributions by the authors to \System{} were made outside their institutional roles. This paper does not claim ownership of the project and should not be read as suggesting institutional endorsement or involvement.

We close by expressing gratitude to all contributors who continue to shape and extend \System{}. We look forward to observing its ongoing evolution.

\section*{Conflict of Interest}
The lead author is the founder of Open WebUI, Inc., which maintains \System{}. This article is an independent academic analysis and was not funded, directed, or otherwise financially supported by Open WebUI, Inc. No confidential or non-public company information was used. The findings and opinions are those of the authors alone and do not represent the views of Open WebUI, Inc. or the authors’ institutions.

\paragraph{Trademarks.} Open WebUI is a mark of Open WebUI, Inc. All other names and marks are the property of their respective owners and are used here for identification purposes only.



\begin{thebibliography}{83}


\ifx \showCODEN    \undefined \def \showCODEN     #1{\unskip}     \fi
\ifx \showDOI      \undefined \def \showDOI       #1{#1}\fi
\ifx \showISBNx    \undefined \def \showISBNx     #1{\unskip}     \fi
\ifx \showISBNxiii \undefined \def \showISBNxiii  #1{\unskip}     \fi
\ifx \showISSN     \undefined \def \showISSN      #1{\unskip}     \fi
\ifx \showLCCN     \undefined \def \showLCCN      #1{\unskip}     \fi
\ifx \shownote     \undefined \def \shownote      #1{#1}          \fi
\ifx \showarticletitle \undefined \def \showarticletitle #1{#1}   \fi
\ifx \showURL      \undefined \def \showURL       {\relax}        \fi
\providecommand\bibfield[2]{#2}
\providecommand\bibinfo[2]{#2}
\providecommand\natexlab[1]{#1}
\providecommand\showeprint[2][]{arXiv:#2}

\bibitem[Any({[n.\,d.]})]%
        {AnythingLLM}
 \bibinfo{year}{[n.\,d.]}\natexlab{}.
\newblock
\newblock
\urldef\tempurl%
\url{https://anythingllm.com/}
\showURL{%
\tempurl}


\bibitem[Lob({[n.\,d.]})]%
        {LobeChat}
 \bibinfo{year}{[n.\,d.]}\natexlab{}.
\newblock
\newblock
\urldef\tempurl%
\url{https://lobechat.com/}
\showURL{%
\tempurl}


\bibitem[Git({[n.\,d.]})]%
        {GitHub}
 \bibinfo{year}{[n.\,d.]}\natexlab{}.
\newblock
\newblock
\urldef\tempurl%
\url{https://www.librechat.ai/}
\showURL{%
\tempurl}


\bibitem[ope({[n.\,d.]})]%
        {openwebui}
 \bibinfo{year}{[n.\,d.]}\natexlab{}.
\newblock \bibinfo{title}{Open WebUI}.
\newblock
\newblock
\urldef\tempurl%
\url{https://github.com/open-webui/open-webui}
\showURL{%
\tempurl}


\bibitem[Ope(2006)]%
        {OpenSourceDefinition2006}
 \bibinfo{year}{2006}\natexlab{}.
\newblock \bibinfo{title}{The {Open} {Source} {Definition}}.
\newblock \bibinfo{howpublished}{\url{https://opensource.org/osd/}}.
\newblock
\urldef\tempurl%
\url{https://opensource.org/osd/}
\showURL{%
\tempurl}


\bibitem[Ste(2023)]%
        {StepsGettingStarted}
 \bibinfo{year}{2023}\natexlab{}.
\newblock \bibinfo{title}{5 {Steps} to {Getting} {Started} with {Llama} 2}.
\newblock
\newblock
\urldef\tempurl%
\url{https://ai.meta.com/blog/5-steps-to-getting-started-with-llama-2/}
\showURL{%
\tempurl}


\bibitem[AIW(2023)]%
        {AIWeightsAre2023}
 \bibinfo{year}{2023}\natexlab{}.
\newblock \bibinfo{title}{{AI} weights are not open "source"}.
\newblock
\newblock
\urldef\tempurl%
\url{https://opencoreventures.com/blog/2023-06-27-ai-weights-are-not-open-source/}
\showURL{%
\tempurl}
\newblock
\shownote{Section: blog}.


\bibitem[Anand et~al\mbox{.}(2023)]%
        {anand2023gpt4all}
\bibfield{author}{\bibinfo{person}{Yuvanesh Anand}, \bibinfo{person}{Zach Nussbaum}, \bibinfo{person}{Adam Treat}, \bibinfo{person}{Aaron Miller}, \bibinfo{person}{Richard Guo}, \bibinfo{person}{Ben Schmidt}, \bibinfo{person}{GPT4All Community}, \bibinfo{person}{Brandon Duderstadt}, {and} \bibinfo{person}{Andriy Mulyar}.} \bibinfo{year}{2023}\natexlab{}.
\newblock \bibinfo{title}{GPT4All: An Ecosystem of Open Source Compressed Language Models}.
\newblock
\newblock
\showeprint[arxiv]{2311.04931}~[cs.CL]


\bibitem[Anthropic(2024)]%
        {claude2024}
\bibfield{author}{\bibinfo{person}{Anthropic}.} \bibinfo{year}{2024}\natexlab{}.
\newblock \bibinfo{title}{Claude}.
\newblock
\newblock
\urldef\tempurl%
\url{https://claude.ai/}
\showURL{%
\tempurl}


\bibitem[Antoniak et~al\mbox{.}(2019)]%
        {antoniak}
\bibfield{author}{\bibinfo{person}{Maria Antoniak}, \bibinfo{person}{David Mimno}, {and} \bibinfo{person}{Karen Levy}.} \bibinfo{year}{2019}\natexlab{}.
\newblock \showarticletitle{Narrative Paths and Negotiation of Power in Birth Stories}.
\newblock \bibinfo{journal}{\emph{Proc. ACM Hum.-Comput. Interact.}} \bibinfo{volume}{3}, \bibinfo{number}{CSCW}, Article \bibinfo{articleno}{88} (\bibinfo{date}{nov} \bibinfo{year}{2019}), \bibinfo{numpages}{27}~pages.
\newblock
\urldef\tempurl%
\url{https://doi.org/10.1145/3359190}
\showDOI{\tempurl}


\bibitem[Arawjo et~al\mbox{.}(2023)]%
        {arawjoChainForgeVisualToolkit2023}
\bibfield{author}{\bibinfo{person}{Ian Arawjo}, \bibinfo{person}{Chelse Swoopes}, \bibinfo{person}{Priyan Vaithilingam}, \bibinfo{person}{Martin Wattenberg}, {and} \bibinfo{person}{Elena Glassman}.} \bibinfo{year}{2023}\natexlab{}.
\newblock \bibinfo{title}{{ChainForge}: {A} {Visual} {Toolkit} for {Prompt} {Engineering} and {LLM} {Hypothesis} {Testing}}.
\newblock
\newblock
\urldef\tempurl%
\url{https://doi.org/10.48550/arXiv.2309.09128}
\showDOI{\tempurl}
\newblock
\shownote{arXiv:2309.09128 [cs]}.


\bibitem[Arawjo et~al\mbox{.}(2024)]%
        {arawjo2024chainforge}
\bibfield{author}{\bibinfo{person}{Ian Arawjo}, \bibinfo{person}{Chelse Swoopes}, \bibinfo{person}{Priyan Vaithilingam}, \bibinfo{person}{Martin Wattenberg}, {and} \bibinfo{person}{Elena~L Glassman}.} \bibinfo{year}{2024}\natexlab{}.
\newblock \showarticletitle{ChainForge: A Visual Toolkit for Prompt Engineering and LLM Hypothesis Testing}. In \bibinfo{booktitle}{\emph{Proceedings of the CHI Conference on Human Factors in Computing Systems}}. \bibinfo{pages}{1--18}.
\newblock


\bibitem[Bai et~al\mbox{.}(2022)]%
        {baiTrainingHelpfulHarmless2022}
\bibfield{author}{\bibinfo{person}{Yuntao Bai}, \bibinfo{person}{Andy Jones}, \bibinfo{person}{Kamal Ndousse}, \bibinfo{person}{Amanda Askell}, \bibinfo{person}{Anna Chen}, \bibinfo{person}{Nova DasSarma}, \bibinfo{person}{Dawn Drain}, \bibinfo{person}{Stanislav Fort}, \bibinfo{person}{Deep Ganguli}, \bibinfo{person}{Tom Henighan}, \bibinfo{person}{Nicholas Joseph}, \bibinfo{person}{Saurav Kadavath}, \bibinfo{person}{Jackson Kernion}, \bibinfo{person}{Tom Conerly}, \bibinfo{person}{Sheer El-Showk}, \bibinfo{person}{Nelson Elhage}, \bibinfo{person}{Zac Hatfield-Dodds}, \bibinfo{person}{Danny Hernandez}, \bibinfo{person}{Tristan Hume}, \bibinfo{person}{Scott Johnston}, \bibinfo{person}{Shauna Kravec}, \bibinfo{person}{Liane Lovitt}, \bibinfo{person}{Neel Nanda}, \bibinfo{person}{Catherine Olsson}, \bibinfo{person}{Dario Amodei}, \bibinfo{person}{Tom Brown}, \bibinfo{person}{Jack Clark}, \bibinfo{person}{Sam McCandlish}, \bibinfo{person}{Chris Olah}, \bibinfo{person}{Ben Mann}, {and} \bibinfo{person}{Jared
  Kaplan}.} \bibinfo{year}{2022}\natexlab{}.
\newblock \bibinfo{title}{Training a {Helpful} and {Harmless} {Assistant} with {Reinforcement} {Learning} from {Human} {Feedback}}.
\newblock
\newblock
\urldef\tempurl%
\url{https://doi.org/10.48550/arXiv.2204.05862}
\showDOI{\tempurl}
\newblock
\shownote{arXiv:2204.05862 [cs]}.


\bibitem[Barr(2023)]%
        {barrMetaMadeIts}
\bibfield{author}{\bibinfo{person}{Alistair Barr}.} \bibinfo{year}{2023}\natexlab{}.
\newblock \bibinfo{title}{Meta made its {Llama} 2 {AI} model open-source because '{Zuck} has balls,' a former top {Facebook} engineer says}.
\newblock
\newblock
\urldef\tempurl%
\url{https://www.businessinsider.com/meta-llama2-open-source-mark-zuckerberg-balls-replit-amjad-masad-2023-10}
\showURL{%
\tempurl}


\bibitem[Baumgartner et~al\mbox{.}(2020)]%
        {baumgartnerPushshiftRedditDataset2020a}
\bibfield{author}{\bibinfo{person}{Jason Baumgartner}, \bibinfo{person}{Savvas Zannettou}, \bibinfo{person}{Brian Keegan}, \bibinfo{person}{Megan Squire}, {and} \bibinfo{person}{Jeremy Blackburn}.} \bibinfo{year}{2020}\natexlab{}.
\newblock \showarticletitle{The {Pushshift} {Reddit} {Dataset}}.
\newblock \bibinfo{journal}{\emph{Proceedings of the International AAAI Conference on Web and Social Media}}  \bibinfo{volume}{14} (\bibinfo{date}{May} \bibinfo{year}{2020}), \bibinfo{pages}{830--839}.
\newblock
\showISSN{2334-0770}
\urldef\tempurl%
\url{https://ojs.aaai.org/index.php/ICWSM/article/view/7347}
\showURL{%
\tempurl}


\bibitem[Benkler et~al\mbox{.}(2015)]%
        {benkler2015peer}
\bibfield{author}{\bibinfo{person}{Yochai Benkler}, \bibinfo{person}{Aaron Shaw}, {and} \bibinfo{person}{Benjamin~Mako Hill}.} \bibinfo{year}{2015}\natexlab{}.
\newblock \showarticletitle{Peer production: A form of collective intelligence}.
\newblock \bibinfo{journal}{\emph{Handbook of collective intelligence}}  \bibinfo{volume}{175} (\bibinfo{year}{2015}).
\newblock


\bibitem[Black et~al\mbox{.}(2022)]%
        {black2022gpt}
\bibfield{author}{\bibinfo{person}{Sid Black}, \bibinfo{person}{Stella Biderman}, \bibinfo{person}{Eric Hallahan}, \bibinfo{person}{Quentin Anthony}, \bibinfo{person}{Leo Gao}, \bibinfo{person}{Laurence Golding}, \bibinfo{person}{Horace He}, \bibinfo{person}{Connor Leahy}, \bibinfo{person}{Kyle McDonell}, \bibinfo{person}{Jason Phang}, {et~al\mbox{.}}} \bibinfo{year}{2022}\natexlab{}.
\newblock \showarticletitle{Gpt-neox-20b: An open-source autoregressive language model}.
\newblock \bibinfo{journal}{\emph{arXiv preprint arXiv:2204.06745}} (\bibinfo{year}{2022}).
\newblock


\bibitem[Blythe and Cairns(2009)]%
        {blythe}
\bibfield{author}{\bibinfo{person}{Mark Blythe} {and} \bibinfo{person}{Paul Cairns}.} \bibinfo{year}{2009}\natexlab{}.
\newblock \showarticletitle{Critical methods and user generated content: the iPhone on YouTube}. In \bibinfo{booktitle}{\emph{Proceedings of the SIGCHI Conference on Human Factors in Computing Systems}} (Boston, MA, USA) \emph{(\bibinfo{series}{CHI '09})}. \bibinfo{publisher}{Association for Computing Machinery}, \bibinfo{address}{New York, NY, USA}, \bibinfo{pages}{1467–1476}.
\newblock
\showISBNx{9781605582467}
\urldef\tempurl%
\url{https://doi.org/10.1145/1518701.1518923}
\showDOI{\tempurl}


\bibitem[Bommasani et~al\mbox{.}(2021)]%
        {bommasani2021opportunities}
\bibfield{author}{\bibinfo{person}{Rishi Bommasani}, \bibinfo{person}{Drew~A Hudson}, \bibinfo{person}{Ehsan Adeli}, \bibinfo{person}{Russ Altman}, \bibinfo{person}{Simran Arora}, \bibinfo{person}{Sydney von Arx}, \bibinfo{person}{Michael~S Bernstein}, \bibinfo{person}{Jeannette Bohg}, \bibinfo{person}{Antoine Bosselut}, \bibinfo{person}{Emma Brunskill}, {et~al\mbox{.}}} \bibinfo{year}{2021}\natexlab{}.
\newblock \showarticletitle{On the opportunities and risks of foundation models}.
\newblock \bibinfo{journal}{\emph{arXiv preprint arXiv:2108.07258}} (\bibinfo{year}{2021}).
\newblock


\bibitem[Candel et~al\mbox{.}(2023)]%
        {candel2023h2ogpt}
\bibfield{author}{\bibinfo{person}{Arno Candel}, \bibinfo{person}{Jon McKinney}, \bibinfo{person}{Philipp Singer}, \bibinfo{person}{Pascal Pfeiffer}, \bibinfo{person}{Maximilian Jeblick}, \bibinfo{person}{Prithvi Prabhu}, \bibinfo{person}{Jeff Gambera}, \bibinfo{person}{Mark Landry}, \bibinfo{person}{Shivam Bansal}, \bibinfo{person}{Ryan Chesler}, \bibinfo{person}{Chun~Ming Lee}, \bibinfo{person}{Marcos~V. Conde}, \bibinfo{person}{Pasha Stetsenko}, \bibinfo{person}{Olivier Grellier}, {and} \bibinfo{person}{SriSatish Ambati}.} \bibinfo{year}{2023}\natexlab{}.
\newblock \bibinfo{title}{h2oGPT: Democratizing Large Language Models}.
\newblock
\newblock
\showeprint[arxiv]{2306.08161}~[cs.CL]


\bibitem[Chen et~al\mbox{.}(2023)]%
        {chen2023phoenix}
\bibfield{author}{\bibinfo{person}{Zhihong Chen}, \bibinfo{person}{Feng Jiang}, \bibinfo{person}{Junying Chen}, \bibinfo{person}{Tiannan Wang}, \bibinfo{person}{Fei Yu}, \bibinfo{person}{Guiming Chen}, \bibinfo{person}{Hongbo Zhang}, \bibinfo{person}{Juhao Liang}, \bibinfo{person}{Chen Zhang}, \bibinfo{person}{Zhiyi Zhang}, \bibinfo{person}{Jianquan Li}, \bibinfo{person}{Xiang Wan}, \bibinfo{person}{Benyou Wang}, {and} \bibinfo{person}{Haizhou Li}.} \bibinfo{year}{2023}\natexlab{}.
\newblock \bibinfo{title}{Phoenix: Democratizing ChatGPT across Languages}.
\newblock
\newblock
\showeprint[arxiv]{2304.10453}~[cs.CL]


\bibitem[Cheng et~al\mbox{.}(2024)]%
        {cheng2024prompt}
\bibfield{author}{\bibinfo{person}{Yu Cheng}, \bibinfo{person}{Jieshan Chen}, \bibinfo{person}{Qing Huang}, \bibinfo{person}{Zhenchang Xing}, \bibinfo{person}{Xiwei Xu}, {and} \bibinfo{person}{Qinghua Lu}.} \bibinfo{year}{2024}\natexlab{}.
\newblock \showarticletitle{Prompt sapper: a LLM-empowered production tool for building AI chains}.
\newblock \bibinfo{journal}{\emph{ACM Transactions on Software Engineering and Methodology}} \bibinfo{volume}{33}, \bibinfo{number}{5} (\bibinfo{year}{2024}), \bibinfo{pages}{1--24}.
\newblock


\bibitem[Contractor et~al\mbox{.}(2022)]%
        {contractor2022behavioral}
\bibfield{author}{\bibinfo{person}{Danish Contractor}, \bibinfo{person}{Daniel McDuff}, \bibinfo{person}{Julia~Katherine Haines}, \bibinfo{person}{Jenny Lee}, \bibinfo{person}{Christopher Hines}, \bibinfo{person}{Brent Hecht}, \bibinfo{person}{Nicholas Vincent}, {and} \bibinfo{person}{Hanlin Li}.} \bibinfo{year}{2022}\natexlab{}.
\newblock \showarticletitle{Behavioral use licensing for responsible AI}. In \bibinfo{booktitle}{\emph{Proceedings of the 2022 ACM Conference on Fairness, Accountability, and Transparency}}. \bibinfo{pages}{778--788}.
\newblock


\bibitem[Depounti et~al\mbox{.}(2023)]%
        {depounti2023ideal}
\bibfield{author}{\bibinfo{person}{Iliana Depounti}, \bibinfo{person}{Paula Saukko}, {and} \bibinfo{person}{Simone Natale}.} \bibinfo{year}{2023}\natexlab{}.
\newblock \showarticletitle{Ideal technologies, ideal women: AI and gender imaginaries in Redditors’ discussions on the Replika bot girlfriend}.
\newblock \bibinfo{journal}{\emph{Media, Culture \& Society}} \bibinfo{volume}{45}, \bibinfo{number}{4} (\bibinfo{year}{2023}), \bibinfo{pages}{720--736}.
\newblock


\bibitem[Eronen and Lee(2024)]%
        {eronen2024improving}
\bibfield{author}{\bibinfo{person}{Juuso Eronen} {and} \bibinfo{person}{Saeun Lee}.} \bibinfo{year}{2024}\natexlab{}.
\newblock \showarticletitle{Improving English Education in Japan: Leveraging Large Language Models for Personalized and Skill-Diverse Learning}.
\newblock  (\bibinfo{year}{2024}).
\newblock


\bibitem[Fan et~al\mbox{.}(2025)]%
        {fan2025can}
\bibfield{author}{\bibinfo{person}{Dongyang Fan}, \bibinfo{person}{Vinko Sabol{\v{c}}ec}, \bibinfo{person}{Matin Ansaripour}, \bibinfo{person}{Ayush~Kumar Tarun}, \bibinfo{person}{Martin Jaggi}, \bibinfo{person}{Antoine Bosselut}, {and} \bibinfo{person}{Imanol Schlag}.} \bibinfo{year}{2025}\natexlab{}.
\newblock \showarticletitle{Can Performant LLMs Be Ethical? Quantifying the Impact of Web Crawling Opt-Outs}.
\newblock \bibinfo{journal}{\emph{arXiv preprint arXiv:2504.06219}} (\bibinfo{year}{2025}).
\newblock


\bibitem[Franzen(2023)]%
        {franzenMetaQuietlyUnveils2023}
\bibfield{author}{\bibinfo{person}{Carl Franzen}.} \bibinfo{year}{2023}\natexlab{}.
\newblock \bibinfo{title}{Meta quietly unveils {Llama} 2 {Long} {AI} that beats {GPT}-3.5 {Turbo} and {Claude} 2 on some tasks}.
\newblock
\newblock
\urldef\tempurl%
\url{https://venturebeat.com/ai/meta-quietly-releases-llama-2-long-ai-that-outperforms-gpt-3-5-and-claude-2-on-some-tasks/}
\showURL{%
\tempurl}


\bibitem[Freelon(2021)]%
        {freelon}
\bibfield{author}{\bibinfo{person}{Deen Freelon}.} \bibinfo{year}{2021}\natexlab{}.
\newblock \showarticletitle{The Post-API Age Reconsidered: Web Science in the ’20s and Beyond}. In \bibinfo{booktitle}{\emph{Proceedings of the 13th ACM Web Science Conference 2021}} (Virtual Event, United Kingdom) \emph{(\bibinfo{series}{WebSci '21})}. \bibinfo{publisher}{Association for Computing Machinery}, \bibinfo{address}{New York, NY, USA}, \bibinfo{pages}{3}.
\newblock
\showISBNx{9781450383301}
\urldef\tempurl%
\url{https://doi.org/10.1145/3447535.3466177}
\showDOI{\tempurl}


\bibitem[Følstad and Skjuve(2019)]%
        {folstadChatbotsCustomerService2019}
\bibfield{author}{\bibinfo{person}{Asbjørn Følstad} {and} \bibinfo{person}{Marita Skjuve}.} \bibinfo{year}{2019}\natexlab{}.
\newblock \showarticletitle{Chatbots for customer service: user experience and motivation}. In \bibinfo{booktitle}{\emph{Proceedings of the 1st {International} {Conference} on {Conversational} {User} {Interfaces}}} \emph{(\bibinfo{series}{{CUI} '19})}. \bibinfo{publisher}{Association for Computing Machinery}, \bibinfo{address}{New York, NY, USA}, \bibinfo{pages}{1--9}.
\newblock
\showISBNx{978-1-4503-7187-2}
\urldef\tempurl%
\url{https://doi.org/10.1145/3342775.3342784}
\showDOI{\tempurl}


\bibitem[Gao et~al\mbox{.}(2020)]%
        {gao2020pile}
\bibfield{author}{\bibinfo{person}{Leo Gao}, \bibinfo{person}{Stella Biderman}, \bibinfo{person}{Sid Black}, \bibinfo{person}{Laurence Golding}, \bibinfo{person}{Travis Hoppe}, \bibinfo{person}{Charles Foster}, \bibinfo{person}{Jason Phang}, \bibinfo{person}{Horace He}, \bibinfo{person}{Anish Thite}, \bibinfo{person}{Noa Nabeshima}, {et~al\mbox{.}}} \bibinfo{year}{2020}\natexlab{}.
\newblock \showarticletitle{The pile: An 800gb dataset of diverse text for language modeling}.
\newblock \bibinfo{journal}{\emph{arXiv preprint arXiv:2101.00027}} (\bibinfo{year}{2020}).
\newblock


\bibitem[Gerganov(2024)]%
        {gerganovGgerganovLlamaCpp2024}
\bibfield{author}{\bibinfo{person}{Georgi Gerganov}.} \bibinfo{year}{2024}\natexlab{}.
\newblock \bibinfo{title}{ggerganov/llama.cpp}.
\newblock
\newblock
\urldef\tempurl%
\url{https://github.com/ggerganov/llama.cpp}
\showURL{%
\tempurl}
\newblock
\shownote{original-date: 2023-03-10T18:58:00Z}.


\bibitem[Goldman(2023)]%
        {goldmanMistralAIBucks2023}
\bibfield{author}{\bibinfo{person}{Sharon Goldman}.} \bibinfo{year}{2023}\natexlab{}.
\newblock \bibinfo{title}{Mistral {AI} bucks release trend by dropping torrent link to new open source {LLM}}.
\newblock
\newblock
\urldef\tempurl%
\url{https://venturebeat.com/ai/mistral-ai-bucks-release-trend-by-dropping-torrent-link-to-new-open-source-llm/}
\showURL{%
\tempurl}


\bibitem[Google(2024)]%
        {bard2024}
\bibfield{author}{\bibinfo{person}{Google}.} \bibinfo{year}{2024}\natexlab{}.
\newblock \bibinfo{title}{Bard - Chat Based AI Tool from Google}.
\newblock
\newblock
\urldef\tempurl%
\url{https://bard.google.com/}
\showURL{%
\tempurl}


\bibitem[Gu et~al\mbox{.}(2021)]%
        {10.1145/3458754}
\bibfield{author}{\bibinfo{person}{Yu Gu}, \bibinfo{person}{Robert Tinn}, \bibinfo{person}{Hao Cheng}, \bibinfo{person}{Michael Lucas}, \bibinfo{person}{Naoto Usuyama}, \bibinfo{person}{Xiaodong Liu}, \bibinfo{person}{Tristan Naumann}, \bibinfo{person}{Jianfeng Gao}, {and} \bibinfo{person}{Hoifung Poon}.} \bibinfo{year}{2021}\natexlab{}.
\newblock \showarticletitle{Domain-Specific Language Model Pretraining for Biomedical Natural Language Processing}.
\newblock \bibinfo{journal}{\emph{ACM Trans. Comput. Healthcare}} \bibinfo{volume}{3}, \bibinfo{number}{1}, Article \bibinfo{articleno}{2} (\bibinfo{date}{oct} \bibinfo{year}{2021}), \bibinfo{numpages}{23}~pages.
\newblock
\urldef\tempurl%
\url{https://doi.org/10.1145/3458754}
\showDOI{\tempurl}


\bibitem[H{\"a}m{\"a}l{\"a}inen et~al\mbox{.}(2023)]%
        {hamalainen2023evaluating}
\bibfield{author}{\bibinfo{person}{Perttu H{\"a}m{\"a}l{\"a}inen}, \bibinfo{person}{Mikke Tavast}, {and} \bibinfo{person}{Anton Kunnari}.} \bibinfo{year}{2023}\natexlab{}.
\newblock \showarticletitle{Evaluating large language models in generating synthetic hci research data: a case study}. In \bibinfo{booktitle}{\emph{Proceedings of the 2023 CHI Conference on Human Factors in Computing Systems}}. \bibinfo{pages}{1--19}.
\newblock


\bibitem[Heikkilä({[n.\,d.]})]%
        {heikkilaFourTrendsThat}
\bibfield{author}{\bibinfo{person}{Heikkilä}.} \bibinfo{year}{[n.\,d.]}\natexlab{}.
\newblock \bibinfo{title}{Four trends that changed {AI} in 2023}.
\newblock
\newblock
\urldef\tempurl%
\url{https://www.technologyreview.com/2023/12/19/1085696/four-trends-that-changed-ai-in-2023/}
\showURL{%
\tempurl}


\bibitem[Hu(2023)]%
        {huChatGPTSetsRecord2023}
\bibfield{author}{\bibinfo{person}{Krystal Hu}.} \bibinfo{year}{2023}\natexlab{}.
\newblock \showarticletitle{{ChatGPT} sets record for fastest-growing user base - analyst note}.
\newblock \bibinfo{journal}{\emph{Reuters}} (\bibinfo{date}{Feb.} \bibinfo{year}{2023}).
\newblock
\urldef\tempurl%
\url{https://www.reuters.com/technology/chatgpt-sets-record-fastest-growing-user-base-analyst-note-2023-02-01/}
\showURL{%
\tempurl}


\bibitem[Ishihara et~al\mbox{.}(2024)]%
        {ishihara2024facilitation}
\bibfield{author}{\bibinfo{person}{Shigekazu Ishihara}, \bibinfo{person}{Taku Ishihara}, {and} \bibinfo{person}{Keiko Ishihara}.} \bibinfo{year}{2024}\natexlab{}.
\newblock \showarticletitle{Facilitation of Kansei Engineering Design Process With LLM Multi-Agent}.
\newblock  (\bibinfo{year}{2024}).
\newblock


\bibitem[Jiang et~al\mbox{.}(2023)]%
        {jiangMistral7B2023}
\bibfield{author}{\bibinfo{person}{Albert~Q. Jiang}, \bibinfo{person}{Alexandre Sablayrolles}, \bibinfo{person}{Arthur Mensch}, \bibinfo{person}{Chris Bamford}, \bibinfo{person}{Devendra~Singh Chaplot}, \bibinfo{person}{Diego de~las Casas}, \bibinfo{person}{Florian Bressand}, \bibinfo{person}{Gianna Lengyel}, \bibinfo{person}{Guillaume Lample}, \bibinfo{person}{Lucile Saulnier}, \bibinfo{person}{Lélio~Renard Lavaud}, \bibinfo{person}{Marie-Anne Lachaux}, \bibinfo{person}{Pierre Stock}, \bibinfo{person}{Teven~Le Scao}, \bibinfo{person}{Thibaut Lavril}, \bibinfo{person}{Thomas Wang}, \bibinfo{person}{Timothée Lacroix}, {and} \bibinfo{person}{William~El Sayed}.} \bibinfo{year}{2023}\natexlab{}.
\newblock \bibinfo{title}{Mistral {7B}}.
\newblock
\newblock
\urldef\tempurl%
\url{https://doi.org/10.48550/arXiv.2310.06825}
\showDOI{\tempurl}
\newblock
\shownote{arXiv:2310.06825 [cs]}.


\bibitem[Jiang et~al\mbox{.}(2024)]%
        {jiang2024mixtral}
\bibfield{author}{\bibinfo{person}{Albert~Q. Jiang}, \bibinfo{person}{Alexandre Sablayrolles}, \bibinfo{person}{Antoine Roux}, \bibinfo{person}{Arthur Mensch}, \bibinfo{person}{Blanche Savary}, \bibinfo{person}{Chris Bamford}, \bibinfo{person}{Devendra~Singh Chaplot}, \bibinfo{person}{Diego de~las Casas}, \bibinfo{person}{Emma~Bou Hanna}, \bibinfo{person}{Florian Bressand}, \bibinfo{person}{Gianna Lengyel}, \bibinfo{person}{Guillaume Bour}, \bibinfo{person}{Guillaume Lample}, \bibinfo{person}{Lélio~Renard Lavaud}, \bibinfo{person}{Lucile Saulnier}, \bibinfo{person}{Marie-Anne Lachaux}, \bibinfo{person}{Pierre Stock}, \bibinfo{person}{Sandeep Subramanian}, \bibinfo{person}{Sophia Yang}, \bibinfo{person}{Szymon Antoniak}, \bibinfo{person}{Teven~Le Scao}, \bibinfo{person}{Théophile Gervet}, \bibinfo{person}{Thibaut Lavril}, \bibinfo{person}{Thomas Wang}, \bibinfo{person}{Timothée Lacroix}, {and} \bibinfo{person}{William~El Sayed}.} \bibinfo{year}{2024}\natexlab{}.
\newblock \bibinfo{title}{Mixtral of Experts}.
\newblock
\newblock
\showeprint[arxiv]{2401.04088}~[cs.LG]


\bibitem[Jiang et~al\mbox{.}(2022)]%
        {jiang2022promptmaker}
\bibfield{author}{\bibinfo{person}{Ellen Jiang}, \bibinfo{person}{Kristen Olson}, \bibinfo{person}{Edwin Toh}, \bibinfo{person}{Alejandra Molina}, \bibinfo{person}{Aaron Donsbach}, \bibinfo{person}{Michael Terry}, {and} \bibinfo{person}{Carrie~J Cai}.} \bibinfo{year}{2022}\natexlab{}.
\newblock \showarticletitle{Promptmaker: Prompt-based prototyping with large language models}. In \bibinfo{booktitle}{\emph{CHI Conference on Human Factors in Computing Systems Extended Abstracts}}. \bibinfo{pages}{1--8}.
\newblock


\bibitem[Jiang et~al\mbox{.}(2019)]%
        {jiang2019moderation}
\bibfield{author}{\bibinfo{person}{Jialun~Aaron Jiang}, \bibinfo{person}{Charles Kiene}, \bibinfo{person}{Skyler Middler}, \bibinfo{person}{Jed~R Brubaker}, {and} \bibinfo{person}{Casey Fiesler}.} \bibinfo{year}{2019}\natexlab{}.
\newblock \showarticletitle{Moderation challenges in voice-based online communities on discord}.
\newblock \bibinfo{journal}{\emph{Proceedings of the ACM on Human-Computer Interaction}} \bibinfo{volume}{3}, \bibinfo{number}{CSCW} (\bibinfo{year}{2019}), \bibinfo{pages}{1--23}.
\newblock


\bibitem[Joa et~al\mbox{.}({[n.\,d.]})]%
        {joadevelopment}
\bibfield{author}{\bibinfo{person}{Byeongha Joa}, \bibinfo{person}{Hanseung Seoa}, \bibinfo{person}{Hyuna Jeonb}, \bibinfo{person}{Joowon Chac}, \bibinfo{person}{Jaejun Leec}, \bibinfo{person}{Seungdon Yeomc}, {and} \bibinfo{person}{Yonggyun Yucd}.} \bibinfo{year}{[n.\,d.]}\natexlab{}.
\newblock \showarticletitle{Development of a Versatile Large Language Model Platform by KAERI: Integrating Intranet and Internet Environments}.
\newblock  (\bibinfo{year}{[n.\,d.]}).
\newblock


\bibitem[Kiene and Hill(2020)]%
        {kieneWhoUsesBots2020}
\bibfield{author}{\bibinfo{person}{Charles Kiene} {and} \bibinfo{person}{Benjamin~Mako Hill}.} \bibinfo{year}{2020}\natexlab{}.
\newblock \showarticletitle{Who {Uses} {Bots}? {A} {Statistical} {Analysis} of {Bot} {Usage} in {Moderation} {Teams}}. In \bibinfo{booktitle}{\emph{Extended {Abstracts} of the 2020 {CHI} {Conference} on {Human} {Factors} in {Computing} {Systems}}} \emph{(\bibinfo{series}{{CHI} {EA} '20})}. \bibinfo{publisher}{Association for Computing Machinery}, \bibinfo{address}{New York, NY, USA}, \bibinfo{pages}{1--8}.
\newblock
\showISBNx{978-1-4503-6819-3}
\urldef\tempurl%
\url{https://doi.org/10.1145/3334480.3382960}
\showDOI{\tempurl}


\bibitem[Kiene et~al\mbox{.}(2019)]%
        {kieneTechnologicalFramesUser2019}
\bibfield{author}{\bibinfo{person}{Charles Kiene}, \bibinfo{person}{Jialun~Aaron Jiang}, {and} \bibinfo{person}{Benjamin~Mako Hill}.} \bibinfo{year}{2019}\natexlab{}.
\newblock \showarticletitle{Technological {Frames} and {User} {Innovation}: {Exploring} {Technological} {Change} in {Community} {Moderation} {Teams}}.
\newblock \bibinfo{journal}{\emph{Proceedings of the ACM on Human-Computer Interaction}} \bibinfo{volume}{3}, \bibinfo{number}{CSCW} (\bibinfo{date}{Nov.} \bibinfo{year}{2019}), \bibinfo{pages}{44:1--44:23}.
\newblock
\urldef\tempurl%
\url{https://doi.org/10.1145/3359146}
\showDOI{\tempurl}


\bibitem[Kovacs et~al\mbox{.}(2018)]%
        {kovacs2018rotating}
\bibfield{author}{\bibinfo{person}{Geza Kovacs}, \bibinfo{person}{Zhengxuan Wu}, {and} \bibinfo{person}{Michael~S Bernstein}.} \bibinfo{year}{2018}\natexlab{}.
\newblock \showarticletitle{Rotating online behavior change interventions increases effectiveness but also increases attrition}.
\newblock \bibinfo{journal}{\emph{Proceedings of the ACM on Human-Computer Interaction}} \bibinfo{volume}{2}, \bibinfo{number}{CSCW} (\bibinfo{year}{2018}), \bibinfo{pages}{1--25}.
\newblock


\bibitem[Kulkarni et~al\mbox{.}(2023)]%
        {kulkarni2023llms}
\bibfield{author}{\bibinfo{person}{Chinmay Kulkarni}, \bibinfo{person}{Tongshuang Wu}, \bibinfo{person}{Kenneth Holstein}, \bibinfo{person}{Q.~Vera Liao}, \bibinfo{person}{Min~Kyung Lee}, \bibinfo{person}{Mina Lee}, {and} \bibinfo{person}{Hariharan Subramonyam}.} \bibinfo{year}{2023}\natexlab{}.
\newblock \showarticletitle{LLMs and the Infrastructure of CSCW}. In \bibinfo{booktitle}{\emph{Companion Publication of the 2023 Conference on Computer Supported Cooperative Work and Social Computing}} (Minneapolis, MN, USA) \emph{(\bibinfo{series}{CSCW '23 Companion})}. \bibinfo{publisher}{Association for Computing Machinery}, \bibinfo{address}{New York, NY, USA}, \bibinfo{pages}{408–410}.
\newblock
\showISBNx{9798400701290}
\urldef\tempurl%
\url{https://doi.org/10.1145/3584931.3608438}
\showDOI{\tempurl}


\bibitem[Liedtke(2023)]%
        {liedtkeGoogleBringsIts2023}
\bibfield{author}{\bibinfo{person}{Michael Liedtke}.} \bibinfo{year}{2023}\natexlab{}.
\newblock \bibinfo{title}{Google brings its {AI} chatbot {Bard} into its inner circle, opening door to {Gmail}, {Maps}, {YouTube}}.
\newblock
\newblock
\urldef\tempurl%
\url{https://apnews.com/article/google-artificial-intelligence-bard-gmail-youtube-maps-1229638b82d19afb5226c913821fa1ad}
\showURL{%
\tempurl}
\newblock
\shownote{Section: Business}.


\bibitem[Lorenz(2023)]%
        {lorenzInfluencerAIClone2023}
\bibfield{author}{\bibinfo{person}{Taylor Lorenz}.} \bibinfo{year}{2023}\natexlab{}.
\newblock \showarticletitle{An influencer’s {AI} clone will be your girlfriend for \$1 a minute}.
\newblock \bibinfo{journal}{\emph{Washington Post}} (\bibinfo{date}{May} \bibinfo{year}{2023}).
\newblock
\showISSN{0190-8286}
\urldef\tempurl%
\url{https://www.washingtonpost.com/technology/2023/05/13/caryn-ai-technology-gpt-4/}
\showURL{%
\tempurl}


\bibitem[Maldeniya et~al\mbox{.}(2020)]%
        {maldeniyaHerdingDelugeGood2020}
\bibfield{author}{\bibinfo{person}{Danaja Maldeniya}, \bibinfo{person}{Ceren Budak}, \bibinfo{person}{Lionel~P. Robert~Jr.}, {and} \bibinfo{person}{Daniel~M. Romero}.} \bibinfo{year}{2020}\natexlab{}.
\newblock \showarticletitle{Herding a {Deluge} of {Good} {Samaritans}: {How} {GitHub} {Projects} {Respond} to {Increased} {Attention}}. In \bibinfo{booktitle}{\emph{Proceedings of {The} {Web} {Conference} 2020}} \emph{(\bibinfo{series}{{WWW} '20})}. \bibinfo{publisher}{Association for Computing Machinery}, \bibinfo{address}{New York, NY, USA}, \bibinfo{pages}{2055--2065}.
\newblock
\showISBNx{978-1-4503-7023-3}
\urldef\tempurl%
\url{https://doi.org/10.1145/3366423.3380272}
\showDOI{\tempurl}


\bibitem[Mesk{\'o} and Topol(2023)]%
        {mesko2023imperative}
\bibfield{author}{\bibinfo{person}{Bertalan Mesk{\'o}} {and} \bibinfo{person}{Eric~J Topol}.} \bibinfo{year}{2023}\natexlab{}.
\newblock \showarticletitle{The imperative for regulatory oversight of large language models (or generative AI) in healthcare}.
\newblock \bibinfo{journal}{\emph{npj Digital Medicine}} \bibinfo{volume}{6}, \bibinfo{number}{1} (\bibinfo{year}{2023}), \bibinfo{pages}{120}.
\newblock


\bibitem[Michael M.~Grynbaum(2023)]%
        {nytimeLawsuit}
\bibfield{author}{\bibinfo{person}{Ryan~Mac Michael M.~Grynbaum}.} \bibinfo{year}{2023}\natexlab{}.
\newblock \bibinfo{title}{The Times Sues OpenAI and Microsoft Over A.I. Use of Copyrighted Work}.
\newblock
\newblock
\urldef\tempurl%
\url{https://www.nytimes.com/2023/12/27/business/media/new-york-times-open-ai-microsoft-lawsuit.html}
\showURL{%
\tempurl}


\bibitem[Morgan(2024)]%
        {morganJmorgancaOllama2024}
\bibfield{author}{\bibinfo{person}{Jeffrey Morgan}.} \bibinfo{year}{2024}\natexlab{}.
\newblock \bibinfo{title}{jmorganca/ollama}.
\newblock
\newblock
\urldef\tempurl%
\url{https://github.com/jmorganca/ollama}
\showURL{%
\tempurl}
\newblock
\shownote{original-date: 2023-06-26T19:39:32Z}.


\bibitem[Nori et~al\mbox{.}(2023)]%
        {nori2023capabilities}
\bibfield{author}{\bibinfo{person}{Harsha Nori}, \bibinfo{person}{Nicholas King}, \bibinfo{person}{Scott~Mayer McKinney}, \bibinfo{person}{Dean Carignan}, {and} \bibinfo{person}{Eric Horvitz}.} \bibinfo{year}{2023}\natexlab{}.
\newblock \bibinfo{title}{Capabilities of GPT-4 on Medical Challenge Problems}.
\newblock \bibinfo{howpublished}{arXiv: 2303.13375}.
\newblock
\urldef\tempurl%
\url{https://www.microsoft.com/en-us/research/publication/capabilities-of-gpt-4-on-medical-challenge-problems/}
\showURL{%
\tempurl}


\bibitem[Nuñez(2023)]%
        {nunezLLaMAHowAccess2023}
\bibfield{author}{\bibinfo{person}{Michael Nuñez}.} \bibinfo{year}{2023}\natexlab{}.
\newblock \bibinfo{title}{{LLaMA} 2: {How} to access and use {Meta}’s versatile open-source chatbot right now}.
\newblock
\newblock
\urldef\tempurl%
\url{https://venturebeat.com/ai/llama-2-how-to-access-and-use-metas-versatile-open-source-chatbot-right-now/}
\showURL{%
\tempurl}


\bibitem[OpenAI(2023)]%
        {openai2023chatgpt}
\bibfield{author}{\bibinfo{person}{OpenAI}.} \bibinfo{year}{2023}\natexlab{}.
\newblock \bibinfo{title}{ChatGPT: {AI} Language Model}.
\newblock \bibinfo{howpublished}{\url{https://openai.com/}}.
\newblock
\newblock
\shownote{Accessed: 2025-08-27}.


\bibitem[OpenAI(2024)]%
        {chatGPT2024}
\bibfield{author}{\bibinfo{person}{OpenAI}.} \bibinfo{year}{2024}\natexlab{}.
\newblock \bibinfo{title}{ChatGPT}.
\newblock
\newblock
\urldef\tempurl%
\url{https://chat.openai.com/}
\showURL{%
\tempurl}


\bibitem[Othman et~al\mbox{.}(2024)]%
        {othman2024comparative}
\bibfield{author}{\bibinfo{person}{Achraf Othman}, \bibinfo{person}{Khansa Chemnad}, \bibinfo{person}{Ahmed Tlili}, \bibinfo{person}{Ting Da}, \bibinfo{person}{Huanhuan Wang}, {and} \bibinfo{person}{Ronghuai Huang}.} \bibinfo{year}{2024}\natexlab{}.
\newblock \showarticletitle{Comparative analysis of GPT-4, Gemini, and Ernie as gloss sign language translators in special education}.
\newblock \bibinfo{journal}{\emph{Discover Global Society}} \bibinfo{volume}{2}, \bibinfo{number}{1} (\bibinfo{year}{2024}), \bibinfo{pages}{1--14}.
\newblock


\bibitem[Park et~al\mbox{.}(2023)]%
        {park2023generative}
\bibfield{author}{\bibinfo{person}{Joon~Sung Park}, \bibinfo{person}{Joseph O'Brien}, \bibinfo{person}{Carrie~Jun Cai}, \bibinfo{person}{Meredith~Ringel Morris}, \bibinfo{person}{Percy Liang}, {and} \bibinfo{person}{Michael~S Bernstein}.} \bibinfo{year}{2023}\natexlab{}.
\newblock \showarticletitle{Generative agents: Interactive simulacra of human behavior}. In \bibinfo{booktitle}{\emph{Proceedings of the 36th Annual ACM Symposium on User Interface Software and Technology}}. \bibinfo{pages}{1--22}.
\newblock


\bibitem[Poudel and Weninger(2024)]%
        {poudel}
\bibfield{author}{\bibinfo{person}{Amrit Poudel} {and} \bibinfo{person}{Tim Weninger}.} \bibinfo{year}{2024}\natexlab{}.
\newblock \showarticletitle{Navigating the Post-API Dilemma}. In \bibinfo{booktitle}{\emph{Proceedings of the ACM Web Conference 2024}} (Singapore, Singapore) \emph{(\bibinfo{series}{WWW '24})}. \bibinfo{publisher}{Association for Computing Machinery}, \bibinfo{address}{New York, NY, USA}, \bibinfo{pages}{2476–2484}.
\newblock
\showISBNx{9798400701719}
\urldef\tempurl%
\url{https://doi.org/10.1145/3589334.3645503}
\showDOI{\tempurl}


\bibitem[Ray(2023)]%
        {ray2023chatgpt}
\bibfield{author}{\bibinfo{person}{Partha~Pratim Ray}.} \bibinfo{year}{2023}\natexlab{}.
\newblock \showarticletitle{ChatGPT: A comprehensive review on background, applications, key challenges, bias, ethics, limitations and future scope}.
\newblock \bibinfo{journal}{\emph{Internet of Things and Cyber-Physical Systems}} (\bibinfo{year}{2023}).
\newblock


\bibitem[Reddit(2024)]%
        {reddit}
\bibfield{author}{\bibinfo{person}{Reddit}.} \bibinfo{year}{2024}\natexlab{}.
\newblock \bibinfo{title}{Upholding {Our} {Public} {Content} {Policy} and {Updating} {Our} robots.txt file - {Upvoted}}.
\newblock
\newblock
\urldef\tempurl%
\url{https://www.redditinc.com/blog/robot-txt-update}
\showURL{%
\tempurl}


\bibitem[Ruan et~al\mbox{.}(2019)]%
        {10.1145/3290605.3300587}
\bibfield{author}{\bibinfo{person}{Sherry Ruan}, \bibinfo{person}{Liwei Jiang}, \bibinfo{person}{Justin Xu}, \bibinfo{person}{Bryce Joe-Kun Tham}, \bibinfo{person}{Zhengneng Qiu}, \bibinfo{person}{Yeshuang Zhu}, \bibinfo{person}{Elizabeth~L. Murnane}, \bibinfo{person}{Emma Brunskill}, {and} \bibinfo{person}{James~A. Landay}.} \bibinfo{year}{2019}\natexlab{}.
\newblock \showarticletitle{QuizBot: A Dialogue-Based Adaptive Learning System for Factual Knowledge}. In \bibinfo{booktitle}{\emph{Proceedings of the 2019 CHI Conference on Human Factors in Computing Systems}} (Glasgow, Scotland Uk) \emph{(\bibinfo{series}{CHI '19})}. \bibinfo{publisher}{Association for Computing Machinery}, \bibinfo{address}{New York, NY, USA}, \bibinfo{pages}{1–13}.
\newblock
\showISBNx{9781450359702}
\urldef\tempurl%
\url{https://doi.org/10.1145/3290605.3300587}
\showDOI{\tempurl}


\bibitem[Schwartz et~al\mbox{.}(2020)]%
        {schwartz2020green}
\bibfield{author}{\bibinfo{person}{Roy Schwartz}, \bibinfo{person}{Jesse Dodge}, \bibinfo{person}{Noah~A Smith}, {and} \bibinfo{person}{Oren Etzioni}.} \bibinfo{year}{2020}\natexlab{}.
\newblock \showarticletitle{Green ai}.
\newblock \bibinfo{journal}{\emph{Commun. ACM}} \bibinfo{volume}{63}, \bibinfo{number}{12} (\bibinfo{year}{2020}), \bibinfo{pages}{54--63}.
\newblock


\bibitem[Sorensen et~al\mbox{.}(2023)]%
        {sorensen2023value}
\bibfield{author}{\bibinfo{person}{Taylor Sorensen}, \bibinfo{person}{Liwei Jiang}, \bibinfo{person}{Jena Hwang}, \bibinfo{person}{Sydney Levine}, \bibinfo{person}{Valentina Pyatkin}, \bibinfo{person}{Peter West}, \bibinfo{person}{Nouha Dziri}, \bibinfo{person}{Ximing Lu}, \bibinfo{person}{Kavel Rao}, \bibinfo{person}{Chandra Bhagavatula}, {et~al\mbox{.}}} \bibinfo{year}{2023}\natexlab{}.
\newblock \showarticletitle{Value Kaleidoscope: Engaging AI with pluralistic human values, rights, and duties}.
\newblock \bibinfo{journal}{\emph{arXiv preprint arXiv:2309.00779}} (\bibinfo{year}{2023}).
\newblock


\bibitem[Studio(2024)]%
        {lmStudio2024}
\bibfield{author}{\bibinfo{person}{LM Studio}.} \bibinfo{year}{2024}\natexlab{}.
\newblock \bibinfo{title}{LM Studio}.
\newblock
\newblock
\urldef\tempurl%
\url{https://lmstudio.ai/}
\showURL{%
\tempurl}


\bibitem[Touvron et~al\mbox{.}(2023)]%
        {touvron2023llama}
\bibfield{author}{\bibinfo{person}{Hugo Touvron}, \bibinfo{person}{Louis Martin}, \bibinfo{person}{Kevin Stone}, \bibinfo{person}{Peter Albert}, \bibinfo{person}{Amjad Almahairi}, \bibinfo{person}{Yasmine Babaei}, \bibinfo{person}{Nikolay Bashlykov}, \bibinfo{person}{Soumya Batra}, \bibinfo{person}{Prajjwal Bhargava}, \bibinfo{person}{Shruti Bhosale}, {et~al\mbox{.}}} \bibinfo{year}{2023}\natexlab{}.
\newblock \showarticletitle{Llama 2: Open foundation and fine-tuned chat models}.
\newblock \bibinfo{journal}{\emph{arXiv preprint arXiv:2307.09288}} (\bibinfo{year}{2023}).
\newblock


\bibitem[Toxtli et~al\mbox{.}(2018)]%
        {10.1145/3173574.3173632}
\bibfield{author}{\bibinfo{person}{Carlos Toxtli}, \bibinfo{person}{Andr\'{e}s Monroy-Hern\'{a}ndez}, {and} \bibinfo{person}{Justin Cranshaw}.} \bibinfo{year}{2018}\natexlab{}.
\newblock \showarticletitle{Understanding Chatbot-Mediated Task Management}. In \bibinfo{booktitle}{\emph{Proceedings of the 2018 CHI Conference on Human Factors in Computing Systems}} (<conf-loc>, <city>Montreal QC</city>, <country>Canada</country>, </conf-loc>) \emph{(\bibinfo{series}{CHI '18})}. \bibinfo{publisher}{Association for Computing Machinery}, \bibinfo{address}{New York, NY, USA}, \bibinfo{pages}{1–6}.
\newblock
\showISBNx{9781450356206}
\urldef\tempurl%
\url{https://doi.org/10.1145/3173574.3173632}
\showDOI{\tempurl}


\bibitem[Tu et~al\mbox{.}(2024)]%
        {tu2024conversational}
\bibfield{author}{\bibinfo{person}{Tao Tu}, \bibinfo{person}{Anil Palepu}, \bibinfo{person}{Mike Schaekermann}, \bibinfo{person}{Khaled Saab}, \bibinfo{person}{Jan Freyberg}, \bibinfo{person}{Ryutaro Tanno}, \bibinfo{person}{Amy Wang}, \bibinfo{person}{Brenna Li}, \bibinfo{person}{Mohamed Amin}, \bibinfo{person}{Nenad Tomasev}, \bibinfo{person}{Shekoofeh Azizi}, \bibinfo{person}{Karan Singhal}, \bibinfo{person}{Yong Cheng}, \bibinfo{person}{Le Hou}, \bibinfo{person}{Albert Webson}, \bibinfo{person}{Kavita Kulkarni}, \bibinfo{person}{S~Sara Mahdavi}, \bibinfo{person}{Christopher Semturs}, \bibinfo{person}{Juraj Gottweis}, \bibinfo{person}{Joelle Barral}, \bibinfo{person}{Katherine Chou}, \bibinfo{person}{Greg~S Corrado}, \bibinfo{person}{Yossi Matias}, \bibinfo{person}{Alan Karthikesalingam}, {and} \bibinfo{person}{Vivek Natarajan}.} \bibinfo{year}{2024}\natexlab{}.
\newblock \bibinfo{title}{Towards Conversational Diagnostic AI}.
\newblock
\newblock
\showeprint[arxiv]{2401.05654}~[cs.AI]


\bibitem[Wang et~al\mbox{.}(2023)]%
        {wang2023enabling}
\bibfield{author}{\bibinfo{person}{Bryan Wang}, \bibinfo{person}{Gang Li}, {and} \bibinfo{person}{Yang Li}.} \bibinfo{year}{2023}\natexlab{}.
\newblock \showarticletitle{Enabling conversational interaction with mobile ui using large language models}. In \bibinfo{booktitle}{\emph{Proceedings of the 2023 CHI Conference on Human Factors in Computing Systems}}. \bibinfo{pages}{1--17}.
\newblock


\bibitem[Wu et~al\mbox{.}(2022a)]%
        {wu2022promptchainer}
\bibfield{author}{\bibinfo{person}{Tongshuang Wu}, \bibinfo{person}{Ellen Jiang}, \bibinfo{person}{Aaron Donsbach}, \bibinfo{person}{Jeff Gray}, \bibinfo{person}{Alejandra Molina}, \bibinfo{person}{Michael Terry}, {and} \bibinfo{person}{Carrie~J Cai}.} \bibinfo{year}{2022}\natexlab{a}.
\newblock \showarticletitle{Promptchainer: Chaining large language model prompts through visual programming}. In \bibinfo{booktitle}{\emph{CHI Conference on Human Factors in Computing Systems Extended Abstracts}}. \bibinfo{pages}{1--10}.
\newblock


\bibitem[Wu et~al\mbox{.}(2022b)]%
        {wu2022ai}
\bibfield{author}{\bibinfo{person}{Tongshuang Wu}, \bibinfo{person}{Michael Terry}, {and} \bibinfo{person}{Carrie~Jun Cai}.} \bibinfo{year}{2022}\natexlab{b}.
\newblock \showarticletitle{Ai chains: Transparent and controllable human-ai interaction by chaining large language model prompts}. In \bibinfo{booktitle}{\emph{Proceedings of the 2022 CHI conference on human factors in computing systems}}. \bibinfo{pages}{1--22}.
\newblock


\bibitem[Wyrich and Bogner(2024)]%
        {wyrich}
\bibfield{author}{\bibinfo{person}{Marvin Wyrich} {and} \bibinfo{person}{Justus Bogner}.} \bibinfo{year}{2024}\natexlab{}.
\newblock \showarticletitle{Beyond Self-Promotion: How Software Engineering Research Is Discussed on LinkedIn}. In \bibinfo{booktitle}{\emph{Proceedings of the 46th International Conference on Software Engineering: Software Engineering in Society}} (Lisbon, Portugal) \emph{(\bibinfo{series}{ICSE-SEIS'24})}. \bibinfo{publisher}{Association for Computing Machinery}, \bibinfo{address}{New York, NY, USA}, \bibinfo{pages}{85–95}.
\newblock
\showISBNx{9798400704994}
\urldef\tempurl%
\url{https://doi.org/10.1145/3639475.3640113}
\showDOI{\tempurl}


\bibitem[Xiao et~al\mbox{.}(2023)]%
        {HEAL}
\bibfield{author}{\bibinfo{person}{Ziang Xiao}, \bibinfo{person}{Wesley~Hanwen Deng}, \bibinfo{person}{Michelle~S. Lam}, \bibinfo{person}{Motahhare Eslami}, \bibinfo{person}{Juho Kim}, \bibinfo{person}{Mina Lee}, {and} \bibinfo{person}{Q.~Vera Liao}.} \bibinfo{year}{2023}\natexlab{}.
\newblock \bibinfo{title}{Workshop on Human Centered Evaluation and Auditing of Large Language Models (HEAL) \@CHI’24}.
\newblock
\newblock
\urldef\tempurl%
\url{https://heal-workshop.github.io/}
\showURL{%
\tempurl}


\bibitem[Xiao et~al\mbox{.}(2020)]%
        {xiaoTellMeYourself2020}
\bibfield{author}{\bibinfo{person}{Ziang Xiao}, \bibinfo{person}{Michelle~X. Zhou}, \bibinfo{person}{Q.~Vera Liao}, \bibinfo{person}{Gloria Mark}, \bibinfo{person}{Changyan Chi}, \bibinfo{person}{Wenxi Chen}, {and} \bibinfo{person}{Huahai Yang}.} \bibinfo{year}{2020}\natexlab{}.
\newblock \showarticletitle{Tell {Me} {About} {Yourself}: {Using} an {AI}-{Powered} {Chatbot} to {Conduct} {Conversational} {Surveys} with {Open}-ended {Questions}}.
\newblock \bibinfo{journal}{\emph{ACM Transactions on Computer-Human Interaction}} \bibinfo{volume}{27}, \bibinfo{number}{3} (\bibinfo{date}{June} \bibinfo{year}{2020}), \bibinfo{pages}{15:1--15:37}.
\newblock
\showISSN{1073-0516}
\urldef\tempurl%
\url{https://doi.org/10.1145/3381804}
\showDOI{\tempurl}


\bibitem[Yablonski(2024)]%
        {yablonski2024laws}
\bibfield{author}{\bibinfo{person}{Jon Yablonski}.} \bibinfo{year}{2024}\natexlab{}.
\newblock \bibinfo{booktitle}{\emph{Laws of UX}}.
\newblock \bibinfo{publisher}{" O'Reilly Media, Inc."}.
\newblock


\bibitem[Zamfirescu-Pereira et~al\mbox{.}(2023a)]%
        {zamfirescu2023johnny}
\bibfield{author}{\bibinfo{person}{JD Zamfirescu-Pereira}, \bibinfo{person}{Richmond~Y Wong}, \bibinfo{person}{Bjoern Hartmann}, {and} \bibinfo{person}{Qian Yang}.} \bibinfo{year}{2023}\natexlab{a}.
\newblock \showarticletitle{Why Johnny can’t prompt: how non-AI experts try (and fail) to design LLM prompts}. In \bibinfo{booktitle}{\emph{Proceedings of the 2023 CHI Conference on Human Factors in Computing Systems}}. \bibinfo{pages}{1--21}.
\newblock


\bibitem[Zamfirescu-Pereira et~al\mbox{.}(2023b)]%
        {10.1145/3544548.3581388}
\bibfield{author}{\bibinfo{person}{J.D. Zamfirescu-Pereira}, \bibinfo{person}{Richmond~Y. Wong}, \bibinfo{person}{Bjoern Hartmann}, {and} \bibinfo{person}{Qian Yang}.} \bibinfo{year}{2023}\natexlab{b}.
\newblock \showarticletitle{Why Johnny Can’t Prompt: How Non-AI Experts Try (and Fail) to Design LLM Prompts}. In \bibinfo{booktitle}{\emph{Proceedings of the 2023 CHI Conference on Human Factors in Computing Systems}} (<conf-loc>, <city>Hamburg</city>, <country>Germany</country>, </conf-loc>) \emph{(\bibinfo{series}{CHI '23})}. \bibinfo{publisher}{Association for Computing Machinery}, \bibinfo{address}{New York, NY, USA}, Article \bibinfo{articleno}{437}, \bibinfo{numpages}{21}~pages.
\newblock
\showISBNx{9781450394215}
\urldef\tempurl%
\url{https://doi.org/10.1145/3544548.3581388}
\showDOI{\tempurl}


\bibitem[Zesch et~al\mbox{.}(2024)]%
        {zesch2024fernuni}
\bibfield{author}{\bibinfo{person}{Torsten Zesch}, \bibinfo{person}{Michael Hanses}, \bibinfo{person}{Niels Seidel}, \bibinfo{person}{Piush Aggarwal}, \bibinfo{person}{Dirk Veiel}, {and} \bibinfo{person}{Claudia de Witt}.} \bibinfo{year}{2024}\natexlab{}.
\newblock \showarticletitle{FernUni LLM Experimental Infrastructure (FLEXI)--Enabling Experimentation and Innovation in Higher Education Through Access to Open Large Language Models}.
\newblock \bibinfo{journal}{\emph{arXiv preprint arXiv:2407.13013}} (\bibinfo{year}{2024}).
\newblock


\bibitem[Zhang et~al\mbox{.}(2020)]%
        {zhangPolicyKitBuildingGovernance2020}
\bibfield{author}{\bibinfo{person}{Amy~X. Zhang}, \bibinfo{person}{Grant Hugh}, {and} \bibinfo{person}{Michael~S. Bernstein}.} \bibinfo{year}{2020}\natexlab{}.
\newblock \showarticletitle{{PolicyKit}: {Building} {Governance} in {Online} {Communities}}.
\newblock In \bibinfo{booktitle}{\emph{Proceedings of the 33rd {Annual} {ACM} {Symposium} on {User} {Interface} {Software} and {Technology}}}. \bibinfo{publisher}{Association for Computing Machinery}, \bibinfo{address}{New York, NY, USA}, \bibinfo{pages}{365--378}.
\newblock
\showISBNx{978-1-4503-7514-6}
\urldef\tempurl%
\url{https://doi.org/10.1145/3379337.3415858}
\showURL{%
\tempurl}


\bibitem[Zhang et~al\mbox{.}(2024)]%
        {zhang2024s}
\bibfield{author}{\bibinfo{person}{Zhiping Zhang}, \bibinfo{person}{Michelle Jia}, \bibinfo{person}{Hao-Ping Lee}, \bibinfo{person}{Bingsheng Yao}, \bibinfo{person}{Sauvik Das}, \bibinfo{person}{Ada Lerner}, \bibinfo{person}{Dakuo Wang}, {and} \bibinfo{person}{Tianshi Li}.} \bibinfo{year}{2024}\natexlab{}.
\newblock \showarticletitle{“It's a Fair Game”, or Is It? Examining How Users Navigate Disclosure Risks and Benefits When Using LLM-Based Conversational Agents}. In \bibinfo{booktitle}{\emph{Proceedings of the CHI Conference on Human Factors in Computing Systems}}. \bibinfo{pages}{1--26}.
\newblock


\bibitem[Zhao et~al\mbox{.}(2023)]%
        {zhao2023survey}
\bibfield{author}{\bibinfo{person}{Wayne~Xin Zhao}, \bibinfo{person}{Kun Zhou}, \bibinfo{person}{Junyi Li}, \bibinfo{person}{Tianyi Tang}, \bibinfo{person}{Xiaolei Wang}, \bibinfo{person}{Yupeng Hou}, \bibinfo{person}{Yingqian Min}, \bibinfo{person}{Beichen Zhang}, \bibinfo{person}{Junjie Zhang}, \bibinfo{person}{Zican Dong}, {et~al\mbox{.}}} \bibinfo{year}{2023}\natexlab{}.
\newblock \showarticletitle{A survey of large language models}.
\newblock \bibinfo{journal}{\emph{arXiv preprint arXiv:2303.18223}} (\bibinfo{year}{2023}).
\newblock


\bibitem[Zheng et~al\mbox{.}(2023)]%
        {zheng2023judging}
\bibfield{author}{\bibinfo{person}{Lianmin Zheng}, \bibinfo{person}{Wei-Lin Chiang}, \bibinfo{person}{Ying Sheng}, \bibinfo{person}{Siyuan Zhuang}, \bibinfo{person}{Zhanghao Wu}, \bibinfo{person}{Yonghao Zhuang}, \bibinfo{person}{Zi Lin}, \bibinfo{person}{Zhuohan Li}, \bibinfo{person}{Dacheng Li}, \bibinfo{person}{Eric~P. Xing}, \bibinfo{person}{Hao Zhang}, \bibinfo{person}{Joseph~E. Gonzalez}, {and} \bibinfo{person}{Ion Stoica}.} \bibinfo{year}{2023}\natexlab{}.
\newblock \bibinfo{title}{Judging LLM-as-a-Judge with MT-Bench and Chatbot Arena}.
\newblock
\newblock
\showeprint[arxiv]{2306.05685}~[cs.CL]
\urldef\tempurl%
\url{https://huggingface.co/spaces/lmsys/chatbot-arena-leaderboard}
\showURL{%
\tempurl}


\end{thebibliography}

\end{document}